\documentclass[12pt]{JHEP3}
\usepackage{graphicx}
\def\a{\alpha}

\def\b{\beta}

\def\cA{{\cal{A}}}
\def\cD{{\cal{D}}}

\def\cL{{\cal{L}}}
\def\cm{{\mathsf{m}}}

\def\cS{{\cal{S}}}

\def\d{\delta}
\def\D{\Delta}

\def\ep{\epsilon}

\def\f{\frac}

\def\G{\Gamma}

\def\k{\kappa}
\def\lf{\left}
\def\l{\lambda}

\def\L{\Lambda}
\def\m{\mu}
\def\n{\nu}

\def\nn{\nonumber}
\def\o{\omega}
\def\O{\Omega}
\def\p{\phi}
\def\P{\Phi}
\def\pa{\partial}

\def\r{\rho}
\def\ra{\rightarrow}

\def\ri{\right}
\def\s{\sigma}

\def\t{\theta}

\def\ti{\tilde}

\def\tr{\textrm}

\def\vp{\varphi}

\def\w{\wedge}
\def\z{\zeta}
\newcommand{\be}{\begin{equation}}
\newcommand{\ee}{\end{equation}}
\newcommand{\bea}{\begin{eqnarray}}
\newcommand{\eea}{\end{eqnarray}}
\newcommand{\barr}{\begin{array}}
\newcommand{\earr}{\end{array}}

\title{Hydrodynamics of R-charged D1-branes}
\author{Justin R. David {${}^a$},Manavendra Mahato {${}^b$},Somyadip
Thakur {${}^a$} and Spenta R. Wadia{${}^{b,c}$} \\
${}^a$ Centre for High Energy Physics,
Indian Institute of Science,\\ C.V. Raman Avenue, Bangalore 560012, India. \\
\email{justin, somyadip@cts.iisc.ernet.in}\\
${}^b$  Department of Theoretical Physics, TIFR,\\
Homi-Bhabha Road,  Mumbai 400005, India.    \\
\email{manav, wadia@theory.tifr.res.in}\\
${}^c$ International Centre for Theoretical Sciences, TIFR, \\
Homi-Bhabha Road,  Mumbai 400005, India.    
}

\abstract{We study the hydrodynamic properties of strongly coupled $SU(N)$
Yang-Mills theory of the D1-brane at finite temperature and at a non-zero 
density of R-charge in the framework of  gauge/gravity duality. 
The gravity dual description involves a charged black hole solution of
an Einstein-Maxwell-dilaton system in 3 dimensions which is obtained by  a
consistent truncation of the spinning D1-brane in 10 dimensions. 
We evaluate  thermal and  electrical conductivity as well as the bulk viscosity
 as a function of  the chemical potential conjugate to the R-charges
 of the D1-brane. We show that the ratio of bulk viscosity to entropy density 
 is independent of the chemical potential and is equal to  $1/4\pi$. 
 The thermal conductivity and bulk viscosity obey a relationship similar 
 to the Wiedemann-Franz law. 
 We show that  at the boundary of thermodynamic stability, the charge 
 diffusion mode becomes unstable and the transport coefficients exhibit critical 
 behaviour. Our method for evaluating the transport coefficients relies on 
 expressing the second order differential equations in terms of a first order equation
 which  dictates the radial evolution 
 of the transport coefficient. The radial evolution equations can be 
 solved exactly for the transport coefficients of our interest. 
 We observe that transport coefficients of the D1-brane theory are related  
 to that of the M2-brane  by an overall proportionality constant which
 sets the dimensions. 
}
\preprint{TIFR-TH/10-24}

\begin{document}

\newpage
\section{Introduction}

There has been recent interest in constructing holographic duals which model phenomena
and properties 
observed in macroscopic low energy physics. 
Such holographic duals may provide 
new insights because the properties and phenomena of interest usually lie in a regime
which is strongly coupled in the field theory description but semi-classical from 
the gravity point of view. 
Transport properties of various systems which admit  holographic duals
have been evaluated from the gravity description. A universal result which
has emerged out of these investigations is that 
the ratio of shear viscosity $\eta$ to the entropy density $s$ for field theories
which admit gravity duals in the two derivative approximation  is given by 
\cite{Kovtun:2003wp,Buchel:2003tz}
 \footnote{See \cite{Rangamani:2009xk} for a recent review and list of 
references on related topics.}
\begin{equation}
\label{sherr}
 \frac{\eta}{s} = \frac{\hbar}{4\pi k_B}, 
\end{equation}
where $\hbar$ is the Planck's constant and $k_B$ is the Boltzmann's constant. 
This ratio has been evaluated for well known $AdS/CFT$ pairs like ${\cal N}=4$ super Yang-Mills 
as well as simple phenomenological gravity models.  Other gauge/gravity duals
which involve  near horizon geometries which are not asymptotically anti-de Sitter
backgrounds like that of Dp-branes, $p\geq 2$  
\cite{Boonstra:1998mp,Itzhaki:1998dd} have also been studied. 
This ratio for these backgrounds has also been shown to be $\hbar/4\pi k_B$
 \cite{Mas:2007ng}.

In \cite{David:2009np}, we began an investigation of macroscopic properties of the 
$1+1$ dimensional field theory of the D1-branes.  
In $1+1$ dimensions, there is no shear, therefore 
it is necessary to study non-conformal field theories to 
obtain non-trivial hydrodynamic coefficients.   
D1-branes are interesting as they 
provide the simplest and the 
most symmetric non-conformal $1+1$ dimensional 
 field theory which admits a gravity dual.
 The theory is the maximally supersymmetric Yang-Mills with $SU(N)$ gauge group. It can be obtained as 
a dimensional reduction of ${\cal N}=4$ SYM from $3+1$ dimensions. 
In \cite{David:2009np}, we  isolated the sound mode in gravity and evaluated the speed of 
sound $v_s$  and the bulk viscosity $\zeta$ in the following regimes
\begin{eqnarray}
 & (i) &\qquad \sqrt{\lambda } N^{-2/3} \ll T\ll \sqrt{\lambda}, \\ \nonumber
& (ii) &\qquad \sqrt{\lambda}N^{-1} \ll T\ll \sqrt{\lambda} N^{-2/3}.
\end{eqnarray}
Here, $\lambda = g_{\rm{YM}}^2 N$ is the t' Hooft coupling and $T$ is the temperature.
In the above regimes, the field theory of the D1-branes admits a gravity dual
\cite{Itzhaki:1998dd} which for the purposes
of evaluating transport coefficients reduces to an Einstein-dilaton theory in $3$ dimensions. 
In \cite{David:2009np}, it was shown that 
\begin{equation}
\label{valtr}
 v_s = \frac{c}{\sqrt{2}}, \qquad \frac{\zeta}{s} = \frac{\hbar}{4\pi k_B}, 
\end{equation}
for hydrodynamics of the D1-brane theory, here 
$c$ is the velocity of light. 
It was also seen that theories arising form D1-branes at cones over
Sasaki-Einstein 7-manifolds  give rise to the values in (\ref{valtr}).
It was suggested that there might be a class of 
non-conformal  field theories  which admit 3d gravitational backgrounds
for which $\zeta/s = \hbar/4\pi k_B$. For the rest of the paper, we will 
work with units in which $\hbar =k_B=c=1$. 

In \cite{Kanitscheider:2008kd}, it was shown that the supergravity fluctuations 
which determine the hydrodynamic coefficients of the 
uncharged D1-branes were related by 
dimensional reduction to that of the M2-branes. The dimensional reduction 
related the shear viscosity  of the conformal hydrodynamics 
of the M2-brane to that of the bulk viscosity of D1-branes. This explained 
why $\zeta/s =1/4\pi$, it can essentially be traced to the relation 
(\ref{sherr}) for the M2-branes. It also explained  why the 
value of the speed of 
sound of the D1-brane theory behaves as though it is a conformal theory in $2+1$ dimensions.
One expects a similar connection for the transport coefficients between the 
D1-brane theory with finite charge density and the corresponding M2-brane 
theory. This would imply that  the ratio $\zeta/s$ will be  
independent of the chemical potential and continues to be
$1/4\pi$ since it is related to the 
ratio $\eta/s$ of the M2-brane theory. 
There  should also be similar relationships between other transport coefficients 
like conductivity.  This is one of our motivations to explore the hydrodynamics
of charged D1-branes.  There is a need to develop novel theories for 1+1 dimensional condensed matter systems as many higher dimensional models can't be applied here and there is a profusion of knowledge  through experiments about new such systems and their properties \cite{Deshpande}. So another reason is to study the macroscopic 
properties of strongly coupled $1+1$ dimensional field theories which 
admit gravitational duals. Gravity duals of $1+1$ dimensional systems 
with a well defined field theory have not been extensively studied 
\footnote{ Holographic duals of 1+1 dimensional systems
from a bottom up approach without a
known boundary field theory  were studied in \cite{Maity:2009zz, Hung:2009qk}.}. These systems play an 
important role in many quantum phenomena and it is worthwhile to 
see what insights the gauge/gravity correspondence gives in this context
with a well defined  field theory in mind. 

In this paper, we study the hydrodynamics of D1-branes 
at finite charge density in a regime which admits a gravity description. 
The gravity dual description involves a charged black hole 
in an Einstein-dilaton-Maxwell scalar system in 3 dimensions which is 
obtained by a consistent truncation of spinning D1-branes in 
10 dimensions. 
We study two situations: 
\begin{enumerate}
\item The case  in which the charge density 
corresponding to a single $U(1)$ of the $SO(8)$ R-symmetry of the 
D1-brane theory is turned on, we call this the 
single charged D1-brane. 
\item 
  The situation in  which equal charge densities
along the $4$ Cartan's are turned on, we call this the equal charged D1-brane.
\end{enumerate}
In both these cases, we see that  both the speed of sound and 
the ratio of bulk viscosity to entropy density is given by (\ref{valtr}). 
The values of these quantities are independent of the chemical 
potential. 
We also evaluate  the charge conductivity, 
 the charge diffusion constant, the sound diffusion
constant
and the thermal conductivity for both the situations and compare 
the results for which 
the corresponding M2-brane calculation has been done. 
We see that apart from 
an overall proportionality constant 
which sets the dimensionality of the 
transport coefficients in these theories,  the transport coefficients
 are identical in the two theories.  The results are summarized in 
the following table. 
 \begin{center}
\begin{tabular}{|c|c|c|c|}
	\hline
Transport        & Single-charged                              & Equal-charged          & Equal-charged         \\
Coefficients     & D1 brane                                    & D1 brane               & M2 brane               \\
	\hline \hline
&&&\\
$\sigma_{DC}$    &$\f{1}{16\pi G_{3}}\f{(2k+3)^2}{9\sqrt{1+k}}$&$\f{1}{16\pi G_{3}}\f{(3-k)^2}{9(1+k)^{2}}$&$\f{1}{16\pi G_{4}}\f{(3-k)^2}{9(1+k)^{2}}$\\
&&&\\
\hline
&&&\\
$\zeta$          &$\f{r_{H}^4}{16 \pi G_{3}L^4}\sqrt{1+k}$ &$\f{r_{H}^4}{16 \pi G_{3}L^4}(1+k)^{2}$&     $--$                     \\
&&&\\
\hline
&&&\\
$\eta$           &          $--$              &         $--$               &$\f{r_{H}^4 (1+k)^2}{16\pi G_{4}L^{\prime 4} }$                           \\ 
&&&\\
\hline
&&&\\
$D_{c}$          &$\f{L^3(3-2k)}{6r_{H}^2\sqrt{1+k}}$&$\f{L^3(k+3)}{6r_{H}^2(1+k)^2}$&                            \\
&&&\\
\hline
&&&\\
$D_{s}$          &$\f{L^3}{12r_{H}^2\sqrt{1+k}}$&$\f{L^3}{12r_{H}^2(1+k)^2}$&     $
\f{L^{\prime 3}} {12r_{H}^2(1+k)^2}$                        \\
&&&\\ 
\hline
&&&\\
$\kappa_T$      & $\frac{r_H^2}{8 L G_3} \frac{(2k+3)(1+k)}{k}$ & $ \frac{r_H^2}{8 L G_3}\frac{(3-k)(1+k)}{k}$
& $ \frac{r_H^2}{8 L' G_4}\frac{(3-k)(1+k)}{k}$ \\
&&&\\ 
\hline
\end{tabular}\\
\vspace {.5cm}
{\bf Table 1. } Transport coefficients of D1-branes and M2-branes.
\end{center}
$r_H$ :  radius of the horizon
$k$: (R-charge)$^2$ in units of $r_H$.\\
$G_3, G_4$: Newton's constant in $3$ and $4$ dimensions. \\
$L, L'$: radius of the orthogonal $S^7$ for D1, M2-branes. \\
$\sigma_{{\rm DC}}$: 
electrical conductivity, $\zeta$: bulk viscosity, $\eta$: shear viscosity. \\
$D_c, D_s$: charge diffusivity, sound diffusivity, $\kappa_T$:
thermal conductivity.

\vspace{1cm}
Hydrodynamics of uncharged M2-branes were first studied in 
\cite{Herzog:2002fn,Herzog:2003ke}. 
We obtained the shear viscosity of the charged M2-branes from the fact that 
 $\eta/s = 1/4\pi$  \cite{Saremi:2006ep}. The 
conductivity  of charged M2-branes 
was obtained from \cite{Hartnoll:2007ip}. The charge diffusion constant
for the M2-branes at non-zero chemical potential has not yet been evaluated
in the literature as far as we are aware. 
 However for M2-branes at zero chemical potential, the charge diffusion 
 constant has been evaluated in \cite{Herzog:2007ij}
 \footnote{See below equation (3.32) in \cite{Herzog:2007ij}.}
 and it agrees with the $k=0$ limit of the 
 D1-brane theory answer. The sound diffusion constant for the 
 charged M2-branes has been calculated by using 
 $D_s = \eta/2(\epsilon +p)$ where $\epsilon, p$ are the energy density and 
 the pressure of the M2-branes.  Notice that the bulk viscosity of the
 D1-brane theory is proportional to the shear viscosity of the 
 M2-brane theory. 
 Another observation of our study of the transport coefficients of the 
 charged D1-brane is the following relationship between the 
 bulk viscosity and the thermal conductivity
 \begin{equation}
   \f{\k _T\hat\mu^2}{\z T}= 4\pi^2 
 \end{equation}
 where $\k _T$ is the thermal conductivity, $T$ the temperature and $\hat \mu$
 the chemical potential. This relationship is analogous  to the 
 Wiedemann-Franz law and a similar relationship between shear 
 viscosity and the thermal conductivity has 
 been observed in the case of single charged D3-branes  \cite{Son:2006em}. 
Since the charged D1-brane theory is obtained as a consistent truncation 
of spinning D1-branes, there is a maximum allowed spin 
or charge beyond which the solution is thermodynamically
unstable \cite{Harmark:1999xt}. 
We show that the transport coefficients exhibit critical behaviour at the boundary 
of the thermodynamical instability. For the single charged case, we observe that 
the charge diffusion mode becomes unstable 
at the boundary of instability. 
This suggests that for this case, the thermodynamical instability can be 
better understood by studying the charged diffusion mode in more detail.

This paper is organized as follows: In the next section, 
we introduce the single charged  D1-brane background and 
obtain the  consistent truncation of the solution to $3$ dimensions. 
We also review the thermodynamics of this solution and obtain the 
boundary of thermodynamic instability. 
 In section 3, we study hydrodynamics of a charged fluid in $1+1$ dimensions 
 and obtain the dispersion relations of the two hydrodynamic  modes, the 
 charge diffusion mode and the sound mode in terms of thermodynamic
 variables. 
 We then use the thermodynamics 
 of the D1-brane solution to explicitly evaluate the dispersion relations. 
 We also determine the form of the retarded correlation functions of the 
 stress tensor and the charge current using conservation laws. 
 In section 4, we study the supergravity fluctuations of the single charged 
 D1-brane solution and isolate the gauge invariant fluctuations  which correspond 
 to the two hydrodynamical  modes in the field theory. 
 In section 5, we determine the various transport coefficients from gravity 
 using the relevant Kubo's formula. To do this, we reduce the problem
 to solving a set of coupled first order non-linear differential equations which are exactly solvable
 in limit required by the Kubo's formulae.  These first order equations dictate the radial
evolution of the transport coefficient. 
 In section 6, we discuss the properties of the transport coefficients, their 
 behaviour at the boundary of thermodynamic instability. We then discuss the
 connection of the D1-brane theory to that  of M2-branes.
It will be interesting to compare our results with what is known
for these systems.
 Appendix A contains the details of the consistent truncation which is required 
 to obtain the charged D1-brane solution in 3 dimensions. Appendix B 
 contains the details of the evaluation of the 
 transport coefficients for the equal charged D1-brane.

\section{The R-charged D1-brane}

In this section, we introduce the gravity dual 
of  $SU(N)$ Yang-Mills with 16 supercharges in $1+1$ dimensions 
at finite R-charge density
 and state its domain of validity. We then discuss its thermodynamic properties. 
This section will also serve to set up notations and conventions. 

In \cite{Itzhaki:1998dd}, it was argued that $SU(N)$ Yang-Mills with $16$ supercharges in $1+1$ dimensions at 
large $N$ is dual to the near horizon geometry of $N$ D1-branes. 
Heating up this theory to a finite temperature $T$, the gravity dual is given in terms of the near horizon 
geometry of non-extremal D1-brane solution.  The gravity dual can be 
 trusted in the domain
\begin{equation}
 \sqrt{\lambda} N^{-\frac{2}{3}} << T << \sqrt{\lambda},
\end{equation}
where $\lambda = \sqrt{g_{\rm YM}^2 N }$ is the t'Hooft coupling of the theory. 
The only non-trivial viscous transport coefficient of this system was evaluated using this gravitational
dual in \cite{David:2009np}. 
We now wish to turn on finite $R$-charge density in the field theory. By the usual gauge/gravity correspondence,
the $SO(8)$ isometry of the $S^7$  present in the near horizon geometry of the 
D1-branes corresponds to the $SO(8)$  R-symmetry of the Yang-Mills. 
Therefore to turn on $R$-charge density, it is necessary to consider D1-branes  with angular momentum.
The near horizon supergravity solution of non-extremal D1-branes  spinning along 
one of the Cartan directions of $SO(8)$ is given by  \cite{Harmark:1999xt}.
\begin{eqnarray}
\label{spinmetric} 
ds^2 &=&  H_1^{-3/4} (  -f dt^2 + dz^2) -2 H_1^{-3/4} \frac{ L^3 r_0^3}{\Delta r^6} l \sin^2 \theta dt d\phi,\nn \\ && + H_1^{1/4} \left( \frac{1}{\tilde h} dr^2  + r^2 ( \Delta d\theta^2 + H \sin^2 \theta d\phi^2 + \cos^2 \theta d\Omega_5^2)
 \right), \nn \\ \nonumber
e^\Phi& =& H_1^{1/2}, \\
A^{(2)} &=&  -\left( \frac{dt}{H_1} + \frac{r_0^3}{L^3} l^2 \sin^2\theta d\phi \right)\wedge dz,
\end{eqnarray}
where
\begin{eqnarray}
 \label{spinmet1}
 \Delta = 1 + \frac{l^2\cos^2\theta}{r^2}, &\qquad& H = 1+ \frac{l^2}{r^2} ,\\ \nonumber 
H_1 = \frac{L^6}{\Delta r^6} , &\quad& f = 1 - \frac{r_0^6}{\Delta r^6}, \\ \nonumber
\tilde h = \frac{1}{\Delta} \left( 1+ \frac{l^2}{r^2} - \frac{r_0^6}{r^6} \right). &\qquad& 
\end{eqnarray}
The above solution  is written in the Einstein frame. 
 $d\Omega_5^2$ is the metric of a unit 5-sphere and 
\begin{equation}
 L^6 = g_{{\rm YM}}^2 2^6 \pi^3 N (\alpha')^4, \qquad
 g_{{ \rm YM}}^2 = \frac{g_s}{2\pi\alpha'}
\end{equation}
with $g_s,\; \alpha'$ being the string coupling and the string length
respectively. 
$A^{(2)}$ is the gauge potential for the RR 2-form sourced by the D1-branes. 
 Note that the above solution reduces to 
the non-spinning  near horizon solution of the non-extremal D1-brane when one sets the 
angular velocity $l=0$.  
For completeness, we mention that the background in (\ref{spinmetric}) is a solution of type IIB supergravity 
equations of motion in 10 dimensions obtained from the following 
action
\begin{equation}
\label{typeiib}
S = \frac{1}{16\pi G_{10} } \int d^{10} x \sqrt{g} \left[  R  - \frac{1}{2}
\partial_M\phi \partial^M\phi - \frac{1}{2 \cdot 3!} e^{\phi} ( F_{3} ) ^2\right].
\end{equation}

To study the hydrodynamics of this solution, one needs to consider perturbations of this solution  
along the brane directions $(t, z)$ and the radial direction. The fluctuations 
along the 7-sphere do not play any role. 
Thus to simplify our analysis, it is
 convenient to perform a Kaluza-Klein reduction 
of this solution  to 3 dimensions.
Using the results of \cite{Cvetic:2000dm},
it can be shown that the 10 dimensional 
solution in (\ref{spinmetric}) admits a  consistent reduction on the $S^7$ sphere to 
 the following solution in 3 dimensions
 \begin{eqnarray}
 \label{finaltruncsol}
ds^2 &=& \left( -c_T^2 dt^2 + c_X^2 dz^2  + c_R^2 dr^2\right) , \\ \nonumber
c_T^2 &=& \left( \frac{r}{L} \right)^8 K, 
\qquad
c_X^2 = \left( \frac{r}{L} \right)^8 H,\qquad c_R^2 = \frac{H}{K} \left( \frac{r}{L}\right)^2,  \\ \nonumber
A_t &=& - \frac{r_0^3 l}{ L^2  ( r^2 + {l^2}) }, \qquad \phi = -3 \log \left( \frac{r}{L} \right) ,\qquad
\Psi = 1 + \frac{l^2}{r^2}.
\end{eqnarray}
Here $H$ and $K$ are defined as
\begin{equation}
\label{defhk}
H = 1 + \frac{l^2}{r^2}, \qquad K = 1 + \frac{l^2}{r^2} - \frac{r_0^6}{r^6}.
\end{equation}
The details of this Kaluza-Klein reduction are given in Appendix A. 
The rotation along one of the Cartan directions reduces to the charge denoted by the 
gauge potential $A_t$ in 3 dimensions. 
Note that the deformation of round $S^7$ metric in (\ref{spinmetric}) 
parametrized by $\Delta$ results in an additional scalar $\Psi$ in 3 dimensions. 
It can also be shown using this consistent reduction that 
the  background in (\ref{finaltruncsol})  is a solution of the equations of
 motion of the following action
 \begin{eqnarray}
 \label{truncact}
I & =& \frac{1}{16\pi G_3} \int d^3 x \sqrt{-g} \left( 
R(g) - \f{8}{9} \partial_\mu \phi \partial^\mu \phi  
-\f{1}{4} \Psi^2 e^{-\frac{4}{3}\phi} F_{\mu\nu} F^{\mu\nu}\ri .\nn\\&&\lf . -\frac{1}{2\Psi^2} \partial_\mu \Psi \partial^\mu \Psi+\f{2}{3\Psi}\pa _{\m}{\p}\pa ^{\m}\Psi  +\f{12}{L^2} e^{\frac{4}{3}\phi} (1+ {\Psi}^{-1})  \right),
\end{eqnarray} 
where
\begin{equation}
\frac{1}{G_{3}} = \frac{ 2\pi^4 L^7} { 3! G_{10}}, \qquad
G_{10} = 2^3 \pi^6 g_s^2 (\alpha')^4.
\end{equation}
Thus the 10 dimensional rotating D1-brane solution reduces to a charged 
black hole of an Einstein-Maxwell-dilaton system along with a scalar. 
The R-charge is given by the gauge potential $A_t$ corresponds to rotation
along the $S^7$ in 10 dimensions.
As a simple consistency check, note that both the action in (\ref{truncact})
and the solution in (\ref{finaltruncsol}) reduces to the truncation studied in 
\cite{David:2009np} \footnote{See equations (4.3), (4.5), (4.6), (4.7).}
 for the uncharged D1-brane. 
Since the above solution is a consistent truncation to 3 dimensions, any solution
to hydrodynamic fluctuations studied in 3 dimensions can by lifted to 10 dimensions.
For completeness, we write down the equations of motion of the action 
given in (\ref{truncact}). 
\begin{eqnarray}
& &G_{\m\n}-\f{1}{2}g_{\m\n}\cA +C_{\m\n}=0,\nn\\
& & \cA=-\f{8}{9}\pa _{\m}\p\pa ^{\m}\p -\f{1}{2\Psi ^2}\pa _{\m}\Psi \pa ^{\m}\Psi+\f{2}{3\Psi}\pa _{\m}\p \pa ^{\m}\Psi-\f{\Psi ^2}{4}e^{-4\p /3}F_{\m\n}F^{\m\n}\\
 & & \qquad\qquad+\f{12}{L^2}e^{4\p /3}(1+\Psi ^{-1}),\\
& & C_{\m\n}=-\f{8}{9}\pa _{\m}\p \pa _{\n}\p-\f{1}{2\Psi ^2}\pa _{\m}\Psi\pa _{\n}\Psi+\f{1}{3\Psi}(\pa _{\m}\p\pa _{\n}\Psi +\pa _{\n}\p\pa _{\m}\Psi )\\
&& \qquad\qquad -\f{1}{2}\Psi ^2e^{-4\p /3}F_{\m\r}{F_{\n}}^{\r},\\
& &\Box \p +\f{6}{L^2}e^{4\p /3}(2+\Psi ^{-1})=0,\\
& & \Box\log\Psi -\f{\Psi^2}{2}e^{-4\p /3}F_{\m\n}F^{\m\n}+\f{8}{L^2}e^{4\p /3}(1-\Psi ^{-1})=0,\\
& &\pa_{\m}[\sqrt{-g}\Psi ^2e^{-4\p /3}F^{\m\n}]=0.
\end{eqnarray}
We refer to the solution in (\ref{finaltruncsol}) as the single charged D1-brane. 
The equal charged D1-brane solution in which equal charge density along
all the $4$ Cartans of the $SO(8)$ are turned on is given in 
(\ref{metEC}) of Appendix A. 

\subsection{Thermodynamics of the R-charged branes}

The thermodynamic properties of spinning D-branes were studied in complete 
generality in 
\cite{Harmark:1999xt} from which we can read out the thermodynamic properties of 
the black hole of interest given in (\ref{finaltruncsol}). We now summarize the relevant 
thermodynamic properties.  The Hawking temperature and the entropy density are given by
\begin{equation}
\label{therm1}
 T = \frac{1}{2\pi L^3} \frac{r_H^5}{r_0^3} ( 3+ 2k), \qquad s = \frac{1}{ 4 G_3} \frac{ r_0^3 r_H}{L^4},
\end{equation}
where $k$  is given by 
\begin{equation}
\label{defkl}
 k = \frac{l^2}{r_H^2},
\end{equation}
 and $r_H$ is the radius of the horizon which is given by the largest root of the  equation
\begin{equation}\label{r0rH}
 r_H^6 + r_H^4 l^2 - r_0^6 =0.
\end{equation}
The energy density and the free energy density is given by
\begin{equation}
\label{therm2}
 \epsilon = \frac{1}{4\pi G_3} \frac{ r_0^6}{L^7}, \qquad p = -f = \frac{1}{8\pi G_3} 
\frac{ r_0^6}{L^7}=  \frac{\epsilon}{2}.
\end{equation}
Here we have also identified the pressure using its thermodynamic relationship with free energy density. 
The charge density $\rho$  and its conjugate the  chemical potential $\mu$  are given by
\begin{equation}
\label{therm3}
 \rho = \frac{1}{8\pi G_3} \frac{ r_0^3 l}{ L^5} ,\qquad \mu = 
 A_{t} (r) |_{r\rightarrow\infty}- A_t(r)|_{r_H}=  \frac{ l r_H^4}{ L^2 r_0^3}.
\end{equation}
Note that we have defined the  chemical potential as the 
 voltage difference between the boundary $r\rightarrow \infty$ and 
 the horizon.  
In writing these thermodynamic quantities, we have used the relation (\ref{r0rH}).  

For the black hole solution given in (\ref{finaltruncsol}) with very large charge, there exists a thermodynamic instability. This instability is equivalent to the instability occurring in D1-branes which are rotating too fast
\cite{Harmark:1999xt}.  Given the energy density of the system, the thermodynamic stability 
is determined by the condition
\begin{equation}
 H_s =  {\rm det} \left( 
 \begin{array}{cc}
 \frac{\partial^2 \epsilon(s, \rho )}{\partial s^2 } & 
 \frac{\partial^2 \epsilon(s, \rho_i)}{\partial s \partial \rho} 
 \\
  \frac{\partial^2 \epsilon(s, \rho_i)}{\partial \rho \partial s}
  & \frac{\partial^2 \epsilon(s, \rho )}{\partial \rho^2 }
  \end{array}\right)
 >0.
\end{equation}
To evaluate it, it is convenient to write the above Hessian as 
\begin{equation}
\label{convhes}
 H_s = \left( \frac{\partial T}{\partial r_0} \frac{\partial \mu}{\partial l } -    \frac{\partial T}{\partial l}
 \frac{\partial \mu}{\partial r_0 } \right)  
\left( \frac{\partial s}{\partial r_0} \frac{\partial \rho}{\partial l } -    \frac{\partial s}{\partial l} \frac{\partial \rho}{\partial r_0 } \right)^{-1} ,
\end{equation}
 where we have used  the chain rule and standard thermodynamic relations. 
Using the expressions for the thermodynamic variables given in (\ref{therm1}), ( \ref{therm2}) and
(\ref{therm3}), it can be shown that the Hessian reduces to 
\begin{equation}
 H_s =  2 G_3^2 L^4 \frac{ ( 3 - 2k)}{ r_H^4 ( 1+k)^2}.
\end{equation}
Thus the condition for thermodynamic stability implies the following restriction on the values of the 
$R$ charge
\begin{equation}
 k< \frac{3}{2}.
\end{equation}
Finally, 
for completeness, we mention that the condition for the validity of the  
supergravity solution of the non-extremal spinning D1-brane
remains the same as that of the non-extremal brane and is given by
\begin{equation}
\label{regime1}
  \sqrt{\lambda} N^{-\frac{2}{3}} << T << \sqrt{\lambda}.
\end{equation}
The bound $k<3/2$ in terms of field theory chemical potential can be written 
as 
\begin{equation}
\label{regime2}
 \hat \mu= \frac{\mu}{ L}  < \frac{\pi T }{\sqrt{6}}.
\end{equation}
Therefore the transport coefficients evaluated in this paper are valid in the 
regime given by (\ref{regime1}) and  (\ref{regime2})  of the field theory.

\section{Hydrodynamics  of a charged fluid in $1+1$ dimensions}

In this section, we show that a charged fluid in $1+1$ dimensions has two hydrodynamic
modes and derive their dispersion relation. 
The stress tensor and the  charge current  of a relativistic fluid in $1+1$ dimensions
are given by 
\begin{eqnarray}
  \label{fluidstress}
T^{\mu\nu} &=& (\epsilon +p) u^{\mu} u^{\nu} + P\eta^{\mu\nu} -\zeta ( u^{\mu} u^\nu + \eta^{\mu\nu})\partial_\lambda u^\lambda, \\ \nonumber
  j^\mu &=& \rho u^\mu -\sigma T (\eta^{\mu\nu}+u^\mu u^\nu)\partial_\nu\left(\frac\mu T\right),
\end{eqnarray}
where $u^\mu$ is the 2-velocity with $u_\mu u^\mu =-1$  and $\zeta$ is the bulk viscosity and 
$\sigma$ the conductivity. The remaining variables $\epsilon, p, \rho, \mu$ 
refer to the energy density, pressure, charge density and the chemical potential of the 
system respectively.   $\eta^{\mu\nu}$ refers to the flat Minkowski metric with 
signature $(-1, 1)$.
The equations of motion of the fluid are given by the following conservation laws
\begin{equation}
\label{conseveq}
\partial_{\mu}T^{\mu\nu} = 0, \qquad\qquad \partial_\mu j^\mu =0.
\end{equation}
We now wish to obtain the linearized hydrodynamics modes, therefore let us
consider small fluctuations from the rest frame of the fluid.
The 2-velocity is then given by
\begin{equation}
 u^0 = 1 , \qquad u^z = \delta u^z.
\end{equation}
Note that $u^0=1$ up to the linear order due to the constraint $u^\mu u_\mu =-1$. 
In considering the small fluctuations, one should keep in mind that spatial  and temporal
variations of the thermodynamic quantities are all of linear order.  
We can write the stress energy tensor to the linear order as given below
\begin{equation}
\label{stress}
 T^{00} = \epsilon + \delta T^{00}, \qquad 
T^{0z} = \delta T^{0z}, \qquad  \delta T^{zz} = p - \frac{\zeta}{\epsilon +p} \partial_z \delta T^{0z}.
\end{equation}
In writing the above form of the stress tensor, 
we have eliminated $\delta u^z$  using 
\begin{equation}
\label{solvux}
  \delta u^z = \frac{ \delta T^{0z}}{ \epsilon +p}, \qquad
\partial_z \delta u^z  = \frac{ \partial_z \delta T^{0z}}{ \epsilon +p}.
\end{equation}
As we are working only to the linear order on 
 taking the spatial derivative of  $\delta u_x$,  the derivative acts only
on $\delta T^{0z}$. This is because 
 derivatives of thermodynamic quantities are first order and therefore contribute 
only  at second order in the above equation. 
Similarly the current density can be written as 
\begin{equation}
\label{current1}
 j^0 = \rho + \delta j^0, \qquad 
j^z = \delta j^z = \rho  \frac{\delta T^{0z}}{ \epsilon +p}  -  \sigma T \partial_z \bar \mu, 
\end{equation}
where $\bar\mu = \mu /T$ and we have again used (\ref{solvux}). 
It is convenient to work with thermodynamic variables in which the 
energy density $\epsilon$ and the charge density $\rho$ are the independent
variables 
and  all other thermodynamic quantities are functions of $\epsilon$ and $\rho$. 
 Then we can 
write $\delta j^z$ as 
\begin{equation}
\label{current2}
 \delta j^z =  \rho  \frac{\delta T^{0z}}{ \epsilon +p} -
 \sigma T \left( \partial_\epsilon \bar \mu \partial_z \delta T^{00} 
 + \partial_{\rho} \bar \mu \partial_z \delta j^0 \right). 
\end{equation}
Substituting the form of the stress tensor and the current density
given in (\ref{stress}), (\ref{current1}) and ( \ref{current2})  into the conservation
equations (\ref{conseveq}), we obtain
\begin{eqnarray}
\label{hydroeq1}
 \partial_0 \delta j^0 +  \rho \frac{\partial_z \delta T^{0z}}{ \epsilon +p}
 -\sigma T\left( \partial_{\epsilon}\bar\mu\partial_{z}^2\ \delta T^{00} +
\partial_{\rho}\bar\mu\partial^2 _{z}\delta j^{0} \right) =0, \\ \nonumber
 \partial_0 \delta T^{00} + \partial_z \delta T^{0z} =0, \\ \nonumber
\partial_0 \delta T^{0z} + \lf (\frac{\partial p}{\partial \epsilon} \partial_z \delta T^{00}+
\frac{\partial p}{\partial\rho}\partial_{z}\delta j^0\ri )
- \frac{\zeta}{\epsilon + p} \partial_z^2 \delta T^{0z}=0.
\end{eqnarray}
The above three equations determine the linearised hydrodynamic modes. 
Performing the Fourier transform of the equations given in (\ref{hydroeq1})
  both in position and 
time, we obtain the following set of algebraic equations 
\begin{eqnarray}
\label{hydroeq2}
( -i\omega + \sigma T\partial_\rho \bar \mu  q^2 ) \delta j^0
+ \frac{ i \rho q}{\epsilon + p} \delta T^{0z} + 
\sigma T \partial_\epsilon \bar \mu q^2  \delta T^{00} =0, \\ \nonumber
- i \omega \delta T^{00} + i q \delta T^{0z} = 0, \\ \nonumber
i q \partial_\epsilon p \delta T^{00} + i q \partial_\rho p \delta j^0 +
\left( - i \omega + \frac{\zeta q^2 }{\epsilon + p} \right) \delta T^{0z} =0.
\end{eqnarray}
The above equations have non-trivial solutions for the fluctuations $\delta j^0, \delta T^{0z}, \delta T^{00}$
only  if the following constraint on $\omega$ is satisfied. 
\begin{eqnarray}
\label{disp1}
 (-i\omega+\sigma Tq^2 \partial_{\rho}\bar\mu)\left(\omega^2- q^2\partial_{\epsilon}p+
\frac{i \zeta q^2\omega }{\epsilon+p}\right)+
q^2\partial_{\rho}p\left(\frac{i\rho \omega}{\epsilon+p}+
\sigma Tq^2\partial_{\epsilon}\bar\mu\right) =0. \nonumber \\
\end{eqnarray}
To solve for $\omega$ in terms of $q$, we can assume the following 
expansions for $\omega$
\begin{equation}
\label{expanom}
\omega =  v_s q - i D_s q^2 + \cdots,  \qquad
\omega = -i D_c q^2+ \cdots . 
\end{equation}
Substituting the first expansion of $\omega$ in terms of 
$q$ given in the above
equation   in the constraint (\ref{disp1}) 
and matching terms of $O(q^3)$ and $O(q^4)$, we 
obtain the following expressions for the sound speed and its damping coefficient
\begin{eqnarray}
 \label{soundspeed}
v_{s}^2 &=&(\partial_\epsilon{p}+\frac{\rho}{\epsilon+p}\partial_\rho{p}), \\ 
\label{sound}
D_s&=&\frac{\zeta}{2(\epsilon+p)}+ \frac{\sigma T}{2v_{s}^2}\lf (\rho\frac{\partial_{\rho}\bar\mu}{\epsilon+p}+
\partial_{\epsilon}\bar\mu \ri )\partial_{\rho}p.
\end{eqnarray}
Similarly 
substituting the second expansion for $\omega$  given in (\ref{expanom}) 
 in the constraint (\ref{disp1}) and demanding that the 
leading coefficient of $O(q^4) $ vanishes, we obtain the following value for 
the charge diffusion constant $D_c$
\begin{equation}
\label{expcdc}
D_c = \sigma T \frac{\partial_{\epsilon}p\partial_{\rho}\bar\mu-\partial_{\rho}p\partial_{\epsilon}\bar\mu}
{\partial_{\epsilon}p+\rho\frac{\partial_{\rho}p}{\epsilon+p}}.
\end{equation}
It can be shown that  these are the only two modes of the equations of motion of linearized hydrodynamics. 
To summarize, the two modes are  the sound mode and the 
charge diffusion mode given by 
the  dispersion relations in ( \ref{expanom}).

We can now use the thermodynamic properties of the charged black hole given in 
(\ref{therm1}), (\ref{therm2}) and (\ref{therm3}) to evaluate the dispersion relations
explicitly. 
From (\ref{therm2}), 
note that the pressure  just depends on the free energy of the system and is independent
of the charge density. Therefore for the R-charged D1-brane, the dispersion relations simplify to 
\begin{eqnarray}
\label{sdiff}
 \omega = \pm \frac{1}{\sqrt{2}}q - i \frac{\zeta}{2(\epsilon +p)}q^2, \\ \nonumber
\omega = - i \sigma T\left.  \frac{\partial \bar \mu }{\partial \rho}\right|_\epsilon q^2.
\end{eqnarray}
We can further simplify the charge diffusion constant as follows
\begin{eqnarray}
\label{Dcval}
 D_c &=& \sigma \left( \partial _\rho \mu - \frac{\mu}{T} \partial _\rho T\right) , \\ \nonumber
&=& \sigma \left( \frac{\partial _l \mu}{\partial _l \rho}  - \frac{\mu}{T} \frac{\partial_l T}{\partial_ l \rho}
\right),
\\ \nonumber
&=&  \s(16\pi G_3)\f{3L^3}{2r_H^2}\f{(3-2k)}{(3+2k)^2}.
\end{eqnarray}
To obtain the second line, we have used chain rule and also the fact that the energy density
$\epsilon$ is independent of $l$.  The last line is obtained by evaluating all the 
the derivatives of the thermodynamic quantities using ( \ref{therm1}), ( \ref{therm2}) and ( \ref{therm3}).
Therefore we see that the charge diffusion mode is given by 
\begin{equation}
\label{cdiff}
\o=-i\s(16\pi G_3)\f{3L^3}{2r_H^2}\f{(3-2k)}{(3+2k)^2}q^2.
\end{equation}
Note that  if the conductivity $\sigma$ remains finite at the boundary of 
thermodynamic stability $k = 3/2$, the charge diffusion mode becomes unstable. 
Later in this paper we will explicitly evaluate the conductivity of the 
charged D1-brane solution and show that it is indeed
finite at $k=3/2$ and thus at the boundary of 
thermodynamic stability, the charge diffusion mode becomes unstable. 

One way of reading out the transport coefficients is to study the hydrodynamic modes and identify the 
coefficient of the dissipative parts.  From (\ref{sdiff}) and (\ref{cdiff}),  we see that we can read out 
both the bulk viscosity and the conductivity. 
Another approach is to use Kubo's formula which directly give the 
transport coefficients in terms of the two point functions. 
Let us first define the various retarded Green's functions:
\begin{eqnarray}
 G_{\mu\nu\alpha\beta}(\omega, q)  &=& -i
\int d^2x \theta(t)e^{ -i(\omega t + q z)} \langle [T_{\mu\nu} (x), T_{\alpha\beta}(0) ]\rangle, 
\\ \nonumber
G_{\mu\nu\rho}(\omega, q)  &=& - i \int d^2 x \theta(t) e^{ -i(\omega t + q z)} 
\langle [J_\mu (x), T_{\nu\rho}(0) ]\rangle, 
\\ \nonumber
G_{\mu\nu}(\omega, q) &=& - i \int d^2 x \theta(t) e^{ -i(\omega t + q z)} 
\langle [ J_\mu(x), J_{\nu} (0)] \rangle.
\end{eqnarray}
Conservation laws and symmetries constrain  the form of 
$ G_{\mu\nu\alpha\beta}(\omega, q)$ to be \cite{David:2009np} 
\begin{equation}
 G_{\mu\nu\alpha\beta}(\omega, q) = P_{\mu\nu} P_{\alpha\beta} G_{B}(\omega, q),
\end{equation}
where $P_{\mu\nu}$ is defined by 
\begin{equation}
 P_{\mu\nu} = \eta_{\mu\nu} - \frac{k_\mu k_\nu} {k^2},
\end{equation}
and  $k_\mu = ( -\omega, q)$.
Thus the two point function of the stress tensor is determined just by 
one function $G_B$.  
For future reference, we write down the following component of this 
correlator
\begin{equation}
G_{zzzz} = \frac{\omega^4}{(\omega^2 - q^2)^2} G_B(\omega, q).
\end{equation}
Similarly one can show that conservation laws $k^\mu G_{\mu\nu}(\omega, q)=0$ 
determine the form of the retarded two point function of the 
currents to be \cite{Kovtun:2005ev}
\begin{equation}
 G_{\mu\nu}(\omega, q) = P_{\mu\nu} G_J(\omega, q).
\end{equation}
We write down the following component of this two point function 
\begin{equation}
G_{zz}(\omega, q) = \frac{\omega^2}{\omega^2 - q^2} G_J(\omega, q).
\end{equation}
What is left now is  the retarded two point function of the stress tensor and the 
charge current. Though we will not be requiring  the form of this two point function, 
for completeness, we state that conservation laws and symmetries determine 
this two point function to be 
\begin{equation}
  G_{\mu\nu\rho}(\omega, q) = \epsilon_{\mu \sigma}k^{\sigma} P_{\nu\rho} G_{S}(\omega, q), 
\end{equation}
where $\epsilon_{\mu\nu}$ is the antisymmetric tensor with $\epsilon_{tz} =-\epsilon_{zt} =1$. 

The transport coefficients, bulk viscosity $\zeta$ and the conductivity $\sigma$ are given 
by the following Kubo's formulae
\begin{eqnarray}
\label{kubo1}
 \zeta = \lim_{\omega\rightarrow 0} \frac{i}{\omega} G_{zzzz}(\omega, q=0)=
 \lim_{\omega\rightarrow 0} \frac{i}{\omega } G_B(\omega, 0),
 \\ \nonumber
 \sigma (\omega) = \frac{i}{\omega} G_{zz}(\omega, q=0)= 
 \frac{i}{\omega} G_J(\omega, 0) .
\end{eqnarray}
The DC conductivity can be obtained by further taking the limit
\begin{equation}
\label{kubo2}
\sigma_{\rm{DC}} =  \lim_{\omega\rightarrow 0}
 \frac{i}{\omega} G_{zz}(\omega, q=0) = \lim_{\omega\rightarrow 0}
 \frac{i}{\omega} G_J(\omega, 0).
\end{equation}
Note that all these formulae involve the $q=0$ limit. 
This is a useful feature which we  will exploit in solving for  the 
hydrodynamic modes from gravity.  We will also be interested in the 
thermal conductivity of the charged D1-brane fluid. The thermal conductivity
can be evaluated using its relation to the electrical conductivity
\cite{Son:2006em}, which is  given by
\begin{equation}
\label{thermcond}
\kappa_T =\left(  \frac{\epsilon + P}{\rho}\right)^2 \frac{\sigma}{T}.
\end{equation}

\section{Hydrodynamic modes in gravity}

In this section, we study linearised fluctuations of the gravity solution in 
(\ref{finaltruncsol}) and isolate the gauge invariant combinations 
of fluctuations which correspond to the sound 
mode and the diffusion mode. These 
 we have obtained in the previous section 
 using general hydrodynamic considerations. 
We consider linearised wave like  perturbations of the single charged D1-brane
solution of the form
$g_{\mu\nu} \rightarrow g_{\mu\nu} + \delta g_{\mu\nu}, 
A_\mu \rightarrow A_{\mu} + \delta A_{\mu}, \phi \rightarrow \phi + \delta\phi$ and
$ \Psi \rightarrow \Psi + \delta\Psi$. 
Due to translational invariance along the D1-brane directions, we can assume
that all the perturbations can be expanded using its Fourier mode as 
\begin{eqnarray}
\delta g_{\mu\nu}(t, z, r ) = e^{ - i( \omega t - q z)} h_{\mu\nu}(r), 
&\qquad& 
 \delta A_{\mu}(t, z, r) = e^{ - i( \omega t - q z)} a_\mu(r), 
\\ \nonumber
\delta\phi (t, z, r) =  e^{ - i( \omega t - q z)} \varphi(r), &\qquad&
\delta \Phi(t, z, r) = e^{ - i( \omega t - q z)} \xi(r).
\end{eqnarray}
We further parameterize the radial dependence of the 
metric and the gauge perturbations as 
\begin{equation}
\label{defpertb}
h_{tt} = - c_T^2  H_{tt}, \qquad 
h_{tz} = c_X^2 H_{tz}, \qquad
h_{zz} = c_X^2 H_{zz}, \qquad
a_\mu = \frac{l^2r_0^3}{L^2} B_\mu, 
\end{equation}
where $c_X$ and $c_T$ are defined in (\ref{finaltruncsol}). 
We fix the gauge by imposing $\delta g_{r\mu} = 0,\; \delta A_r =0$. 
The linearized equations of motion for the perturbations are given by
\begin{eqnarray}
&&3r^2H^2KH_{zz}''+3rH\lf [\lf (2H+1\ri )H+\lf (3H+1\ri )K\ri ]H_{zz}'+6Kr\xi '\nn\\
&& -4HK\lf (3H+1\ri )r\vp '+{6}{r}^3H^2(H-1)(H-K)B_t'+12\lf (4H-1\ri )\xi\nn \\ \label{EqHzz}
&& -6(H-1)(H-K)H_{tt}-8\lf \{6H^2\lf (H+1\ri )+(H-1)(H-K)\ri \}\vp=0, \\
&& rHH_{tz}''+(5H+2)H_{tz}'+2r^2(H-1)(H-K)B_z'=0, \label{Exp12}\\
&& 3r^2H^2KH_{tt}''+3rH\lf [3H\lf (2H+1\ri )-\lf (H+1\ri )K\ri ]H_{tt}'+6Kr\xi '\nn\\
&& -4HK\lf (3H+1\ri )r\vp '-{6}{r}^3H^2(H-1)(H-K)B_t'+12\lf (2H+1+\f{2l^2}{r^2}\f{K}{H}\ri )\xi\nn\\ \label{EqHtt}
&& +6(H-1)(H-K)H_{tt}-8\lf \{6H^2\lf (H+1\ri )-(H-1)(H-K)\ri \}\vp=0, \\
&& H^3B_t''+\f{H^2}{r}\lf (4-H\ri )B_t'+\f{H}{r^3}\lf (\f{4}{H}\xi '-\f{8}{3}\vp '+H_{zz}'-H_{tt}'\ri )\label{New1}\nn\\
&&\qquad +\f{8}{H}\f{(H-1)}{r^4}\xi -\f{L^6}{r^6}\f{H^3}{K}q(\o B_z+qB_t)=0, \\
&& HKB_z''+\f{1}{r}\lf \{2H\lf (2H+1\ri )-\lf (5H-2\ri )K\ri \}B_z'+\f{2}{r^3}H_{tz}'  \nn\\
&& \qquad\qquad+\f{L^6}{r^6}\f{H^2}{K}\o (q B_t+\o B_z)=0, \label{New2}\\
&&r^2\vp ''+\lf [1+\f{2}{K}\lf (2H+1\ri )\ri ]r\vp '-\f{3}{2}r(H_{tt}+H_{zz})'\nn\\ \label{Eqphi}
&& -\f{6}{KH}\xi +\f{1}{K^2}\lf [8K\lf (2H+1\ri )+\f{L^6}{r^4}(\o ^2H-q^2K)\ri ]\vp=0, \\
&&  r^2HK\xi ''+\lf [2H\lf (2H+1\ri )+K\lf (5H-4\ri )\ri ]r\xi '-\f{l^2}{r}HK(H_{zz}+H_{tt})' \nn\\
&& +4r^3H^2(H-1)(H-K)B_t'+\f{16}{3}\f{l^2}{r^2}\lf (2H^2+K-H\ri )\vp -4(H-1)(H-K)H_{tt}\nn\\ 
&& +\lf [(\o ^2H-q^2K)\f{H}{K}\f{L^6}{r^4}+4\lf \{H(2H+3)-3-\lf (4-H\ri )(H-1)\f{K}{H}\ri \}\ri ]\xi
= 0.\label{Eqxi}
\end{eqnarray}
Here $H$ and $K$ are defined in (\ref{defhk}).
Equations of motion obtained from the variations $\delta g_{\mu r}$ and $\delta A_{r}$ lead to 
the following $4$ constraints.
\begin{eqnarray}
&& rH(qKH_{tt}'-\o HH_{tz}')+q\lf (2H+1\ri )(H-K)H_{tt}
-\f{4}{3}qK(3H+1)\vp +2q\f{K}{H}\xi
 \nn\\ \label{Con1}
&&\qquad\qquad -2r^2H(H-1)(H-K)(qB_t+\o B_z)=0, \\
&& rH^2(qH_{tz}+\o H_{zz})'+2\o \xi-\f{4}{3}\o H(3H+1) \vp \nn\\
&& \qquad\qquad -\f{H}{K}(H-K)\lf (2H+1\ri )(\o H_{zz}+2qH_{tz})=0, \label{Con2}\\
&& 3rH^2K\lf (3H+1\ri )H_{tt}'+3rH^3\lf (K+2H+1\ri )H_{zz}'+4rH^2K\lf (3H+1\ri )\vp '  -6rHK\xi '\nn\\
&& +6r^3H^3(H-1)(H-K)B_t' +12\lf \{\lf (2H+1\ri )H+2(H-1)(H-K)\ri \}\xi \nn\\
&& -8H\lf \{6H^2(H+1)+(H-1)(H-K)\ri \}\vp -6H(H-1)(H-K)H_{tt} \nn \\ 
&&+3\f{H^4}{K}\f{L^6}{r^4}\lf(-q^2 \f{K}{H}H_{tt}+2\o q H_{tz}+\o ^2H_{zz}\ri )  =0,
 \label{Con3}\\
&& r^3H^2\lf (\o B_t'+q\f{K}{H}B_z'\ri )+2qH_{tz}-\o \lf (H_{tt}-H_{zz}+\f{8}{3}\vp-\f{4}{H}\xi\ri )=0. \label{New3}
\end{eqnarray}
It can be shown that the constraints 
(\ref{Con1}), ( \ref{Con2}), ( \ref{Con3}) and ( \ref{New3}) 
are consistent with the dynamical equations of motion  (\ref{EqHzz}), 
( \ref{Exp12}), ( \ref{EqHtt}), (\ref{New1}),  ( \ref{New2}), ( \ref{Eqphi}) and 
(\ref{Eqxi}).  That is, on evolving the constraints using the equations of motion, one does not 
generate new constraints.
We have verified  that on differentiating the constraints with 
respect to $r$, one just obtains a linear combination of the dynamical equations of motion 
as well as the constraints. 

Though we have fixed the gauge $\delta g_{\mu r} =0,\; \delta A_r =0$, there are 
still residual gauge degrees of freedom arising from diffeomorphisms
$x^\mu \rightarrow x^\mu + \epsilon^\mu $ with 
$\epsilon^\mu = \epsilon^\mu(r, \omega, q) e^{-i\omega t + i q z}$ 
and $U(1)$ gauge transformations
$A_\mu \rightarrow A_\mu + \partial_\mu \chi$ with 
$\chi$ = $\tilde\chi(\omega, q)$ $ e^{-i\omega t + i q z}$.
Under diffeomorphism, the metric, 
the gauge field and the scalars transform as 
\begin{eqnarray}
 g_{\mu\nu} &\rightarrow& g_{\mu\nu}  - \nabla_\mu \epsilon_\nu - 
\nabla_\nu \epsilon_\mu, \\ \nonumber
A_\mu &\rightarrow& A_\mu  - \partial_\mu \epsilon^\rho A_\rho -
 \epsilon^\sigma\partial_\sigma A_\mu , \\ \nonumber
\phi &\rightarrow&  \phi -\partial_\mu \phi \epsilon^\mu, \qquad
\Psi \rightarrow  \Psi -\partial_\mu \Psi \epsilon^\mu, 
\end{eqnarray}
where 
$\epsilon^\mu(r, \omega, q)$ is determined  by the  gauge condition 
$\delta g_{\mu r} =0$. 
The residual $U(1)$ gauge transformations on  a given 
Fourier mode of the gauge field act as follows
\begin{equation}
 A_t \rightarrow A_t    -i\omega \tilde\chi, \qquad A_z \rightarrow A_z + i q \tilde \chi.
\end{equation}
Instead of fixing the gauge completely, it is more convenient to work in 
variables which are invariant under these residual gauge transformations. 
To do this, we first work out the change of the fluctuations under diffeomorphisms
explicitly. This is given by
\begin{eqnarray}
\label{diffeot}
 H_{tt}\ra H_{tt}  -\f{2}{c_T^2}(i\o \epsilon _t+\G ^r_{tt}\epsilon _r), &\qquad &\nn\\
H_{tz}\ra H_{tz}+ \f{1}{c_X^2}(i\o \epsilon _z-iq\epsilon _t), &\qquad& 
H_{zz}\ra H_{zz} -\f{2}{c_X^2}(iq\epsilon _z-\G ^r_{zz}\epsilon _r).\nn\\
B_t\ra B_t-\f{2}{r^3H^2}\epsilon ^r+i\o \f{L^2A_t}{lr_0^3}\epsilon ^t, &\qquad& 
B_z\ra B_z-iq\f{L^2A_t}{lr_0^3}\epsilon ^t. \nn\\
\vp \ra \vp -\f{\p '}{c_R^2}\epsilon _r,  &\qquad& 
\xi \ra \xi -\f{H'}{c_R^2}\epsilon _r ,
\end{eqnarray}
where $\Gamma$'s refer to the Christoffel symbols of the single charged D1-brane
solution. 
Similarly under the $U(1)$ transformations, the gauge field fluctuations change as
\begin{equation}
\label{gauget}
 B_t \rightarrow B_t   -i \omega \tilde \chi, \qquad B_z \rightarrow B_z + i q \tilde\chi.
\end{equation}
From the gauge transformations in (\ref{diffeot}) and ( \ref{gauget}), we can show that 
the following are gauge invariant variables both under diffeomorphisms as 
well as $U(1)$ gauge transformations. 
\begin{eqnarray}
\label{gainv}
Z_P&=&-q^2\f{K}{H}H_{tt}+2\o qH_{tz}+\o ^2H_{zz}-\f{2V}{3H}\vp, \nn\\
G_P&=&qB_t+\o B_z+\f{2q}{3r^2H^2}\vp, \nn\\
S_P&=&2(1-H)\vp +3\xi .
\end{eqnarray}
where
\begin{eqnarray}
V&=&q^2(K+2H+1)-\o ^2(3H+1)
\end{eqnarray}
and $k$ is defined in (\ref{defkl}).
Note that the gauge invariant 
variables given in (\ref{gainv}) 
are not unique, in fact any  linear combinations of the above  variables are 
also gauge invariant. 

After tedious but straightforward manipulations, it can be shown that the 
dynamical equations and the constraint equations can be used to 
write down $3$ second order coupled linear differential equations for the 
gauge invariant variables $Z_P, G_P$ and $ S_P$. 
Before we present these equations, we redefine quantities so that 
we are dealing only with dimensionless variables as follows:
\begin{eqnarray}
\label{dimension}
\f{r_H^2}{r^2}&=& u, \qquad
\hat{q}=r_H q, \qquad
\hat{\o} =r_H \o,  \nn\\
\hat{Z}_P&=&r_H^2 Z_p, \qquad
\hat{G}_P=r_H^{3}G_P, \qquad
\hat{S}_P=S_P\nn\\
r_0^6&=&r_H^6( 1+ k)\quad \mbox{with}\quad  k= \frac{l^2}{r_H^2}, \qquad
\hat{L}=\f{L}{r_H}.
\end{eqnarray}
We also define the expression
\begin{equation}
\a _t=q^2\f{K}{H}-\o ^2.
\end{equation}
In the equations below, for brevity of notation, we continue to refer to the hatted
dimensionless quantities in terms of their original symbols.
The equations for the gauge invariant quantities  are given below where 
the prime denotes derivative with respect to the dimensionless quantity
$u$.
\bea
\label{zpu}
Z_P''&+&\lf [\f{(K-2H-1)}{uK}+\f{2}{uHV}\{q^2(H-K)(2H+1)+(H-1)(q^2-\o ^2)\}\ri ]Z_P'\nn\\&=&
\lf [\f{L^6H}{4K^2}\a _t+\f{(H-K)}{u^2HVK}\{q^2(K(4H+5)-(2H+1)^2)-\o ^2(H-1)\}\ri ]Z_P\nn\\&&
+\f{2q(H-K)(H-1)}{u^3H V K} [q^2\{K(4H+5)-(2H+1)^2\}\nn\\&&
-\o ^2(H-1)+6\o ^2(H+1)(H-K) ]G_P\nn\\&&
+\f{1}{3u^2VH^3K}\lf [-q^4(H-K)^2(3H+10)+6q^2\o ^2(H-K)^2(H+2)\ri.\nn\\&&
\lf . +(9H^2-2H+1)H(q^2-\o ^2)^2-6H(H-1)(H-K)(q^2-\o ^2)^2\ri.\nn\\&&
\lf. -2(H-K)q^2(q^2-\o ^2)(2H^2-H+1)\ri ]S_P,
\eea
\bea\label{gpu}
G_P''&-&\f{G_P'}{uH^2K\a _t}[2q^2K^2(1-H)+\o ^2H\{K(4H-1)-H(2H+1)\}]\nn\\&&
+\f{q}{H^2V\a _t}[q^2(2H-K+1)-\o ^2(H+1)]Z_P'-\f{2q}{3H^3}S_P'\nn\\&=&
\f{G_P}{VH^3K^2\a_t}\lf[\f{L^6}{4}H^4V\a _t^2+\f{K}{u^2}(H-1)(H-K)\{q^2(2H+1)(2q^2K\ri. \nn\\&&\lf.
-\o ^2(H+K))
-\o ^2\a _tH(3H+1) \}\ri]+\f{q}{uH^2VK}(2H+1)(H-K)Z_P\nn\\&&
+\f{q}{3uH^5KV\a _t}[(1+3H-6H^2)H^2(q^2-\o ^2)^2\nn\\&&
+q^4(H-K)\{2H+(3H+2)HK+2(H-1)K^2\}+3q^2HV(H-1)(H-K)\nn\\&&
+\o ^2H(H-K)\{q^2(H^2-6KH-2)+2H\a _t+2V(H+1)\}]S_P,
\eea
\bea\label{spu}
S_P''&-&\f{6(H-1)(q^2-\o ^2)}{uV\a _t}Z_P'-\f{6q}{u^2\a _t}(H-1)(H-K)G_P'\nn\\&&
-\f{1}{u H K}[H(2H+1)+K(H-2)]S_P'=\nn\\&&
-\f{3}{u^2V K}(H-1)(H-K)Z_P-\f{6q}{u^3V H \a _t}(H-1)^2(H-K)(q^2-\o ^2)G_P\nn\\&&
+\f{1}{u^2H^3KV\a _t}\lf[\f{L^6u^2}{4K}VH^4\a _t^2+(4-H)(H-1){H^2K}\a _t^2
\ri.\nn\\&&
-(q^2-\o ^2)^2H^2(8H^3+5H^2-7H+2)\nn\\&&
+q^2H(H-K)\{\a _t(H+1)(4H^2-9H+6)+q^2(8H^3+5H^2-7H+2)\}\nn\\&&
 +\o ^2(H-K)\{q^2H(2+11H-13H^2-8H^3)-H(H-1)\a _t(2H^2-3H-6)\nn\\&&
\lf.+4\o ^2H(H-1)(2H+1)\}\ri]S_P.
\eea
At present, it seems that there are $3$ gauge invariant modes in contrast with the 
$2$ modes  in  $1+1$ hydrodynamics  as shown in the previous section. 
We will show subsequently that one of these modes can be decoupled from the 
rest and consistently set to zero and plays no role in determining the 
transport coefficients. 

\subsection{Properties of the fundamental equations}

Though the equations given in (\ref{zpu}), ( \ref{gpu}) and ( \ref{spu}) seem
a set of complicated coupled differential equations, we will show that for
the transport properties of interest, namely the conductivity and the bulk viscosity
can be obtained from them using analytical methods. 
For this purpose, we need to discuss various properties relevant 
to these equations. 

\vspace{.5cm}
\noindent
{\bf {(i) $l =0$ limit}}
\vspace{.5cm}

In this limit, the charged D1-brane reduces to the uncharged 
D1-brane. 
An important check for the system of equations in 
(\ref{zpu}), ( \ref{gpu}) and ( \ref{spu}) is that they decouple and one of the mode
reduces to the sound mode studied in \cite{David:2009np}. 
Setting $l=0$, we  see  the parameters which enter these equations 
reduce to 
\begin{eqnarray}
r_0&\rightarrow&r_H, \qquad
 H \rightarrow  1, \qquad K \rightarrow f = 1 - \frac{r_0^6}{r^6},  \\ \nonumber
\a _t&\rightarrow&q^2(f-\l), \nn\\
V&\rightarrow&q^2(3+f-4\l), \qquad \mbox{with} \quad \lambda = \frac{\omega^2}{q^2}.
\end{eqnarray}
Note that with these parameters, the mode $Z_P$ reduces to the 
sound mode studied in $\cite{David:2009np}$,  also 
the definition of  $H_{tt}$ is negative of $H_{tt}$ in 
\cite{David:2009np}.  
Substituting these values of the parameters into the fundamental equations
for the gauge invariant fluctuations, we see that the variable $S_P$ can be consistently
set to zero and the equation for $Z_P$  decouples from $G_P$.
The equation 
for $Z_P$ reduces to 
\begin{equation}
Z_P''+\lf[\f{6+f}{rf}-\f{12(1-f)}{r(3+f-4\l)}\ri ]Z_P'-\lf [\f{q^2L^6}{f^2r^6}(f-\l)-\f{36(1-f)^2}{r^2f(f+3-4\l)}\ri ]Z_P=0.
\end{equation}
It can be seen that this is the equation for the sound mode obtained in 
\cite{David:2009np}. 
To write the equation for the gauge fluctuation
in the $l=0$ limit,  it is convenient to redefine it as 
\begin{eqnarray}
H_P = \f{lr_0^3}{L^2}G_P&=&
q\lf (\f{lr_0^3}{L^2}B_t\ri )+\o \lf (\f{lr_0^3}{L^2}B_z\ri )+\f{2qlr_0^3}{3r^2H^2L^2}\vp, \nn\\
&=&q \delta A_t+\o \delta A_z+\f{2qlr_0^3}{3r^2H^2L^2}\vp.
\end{eqnarray}
Thus in the $l\rightarrow 0$ limit, the dilaton fluctuation decouples from the 
gauge invariant combination $H_P$. 
Substituting $G_P$ in terms of $H_P$, it can be seen that the 
sound mode $Z_P$ decouples from the gauge mode and reduces to 
\begin{equation}
\label{fineqnH}
 H_P''-\frac{3u^2\lambda}{(\lambda-f)f}H_P'+
 \frac{L^3}{4f^2} (\omega^2-q^2 f)H_P=0.
\end{equation}
It can be easily verified that this is the equation which is obtained 
by examining  the gauge field equation
\begin{equation}
\partial_\mu ( \sqrt{-g} e^{-4\phi/3} F^{\mu\nu} ) =0,
\end{equation}
where the metric  and the dilaton background values are that of the uncharged 
D1-brane. The background gauge field in this case vanishes and the 
field strength $F^{\mu\nu}$ is just that of the fluctuations $\delta A_z$ and $ \delta A_t$. 
Thus we have seen that in the $l=0$ limit, we obtain two modes, the mode $Z_P$ 
corresponds to the sound mode and the mode $H_P$ 
corresponds to the charge diffusion mode. 
This is what is expected for the uncharged D1-brane. 
 The dispersion relation for  the quasi-normal mode of 
 $Z_P$ 
 was obtained in \cite{David:2009np} and it is given by
 \be\label{dispersionrelg}
\o = \pm\frac{1}{\sqrt{2}} q - i \frac{L^3}{12}q^2+....
\ee
Note that we are measuring all quantities in units of $r_H=r_0$ here. 
Then identifying the sound speed and the bulk viscosity 
from the above dispersion relation, it was seen that 
\begin{equation}
\label{speedvisco}
 v_s^2 =\frac{1}{2}, \qquad  \frac{\zeta}{s} =  \frac{1}{4\pi},
\end{equation}
where $s$ represents entropy density for the uncharged D1-brane. 
In this paper, we will show that the ratio $\zeta/s$ continues to be
$\frac{1}{4\pi}$ for the case of the charged D1-brane also.
The quasi-normal mode  for  the gauge field equation  (\ref{fineqnH}) 
 is given by 
\begin{eqnarray}
 H_P &=& A ( 1-u^3)^{-i\frac{L^3}{6} \omega} \left( 
1 + i \omega \frac{L^3}{2} \left[ \frac{1}{2}  \ln  \frac{1+ u + u^2}{3}    \right. \right.  \\ \nonumber
 & & \left. \left.  +
 \frac{1}{\sqrt{3}}\left\{ \tan^{-1}\left( \frac{2u+1}{\sqrt{3} }\right) - \frac{\pi}{3}\right\} \right]
 +  i \frac{q^2 L^3}{2\omega} ( 1-u)  + O( \omega^2, q^4,  \omega q^2) 
\right) 
\end{eqnarray}
where $A$ is an arbitrary constant. 
Note that the above solution satisfies the ingoing boundary condition at the horizon $u=1$. Imposing 
the Dirichlet condition  at the boundary $u=0$, we obtain the charge dispersion relation 
\begin{equation}
 \omega = - i \frac{L^3}{2} q^2 + \cdots. 
\end{equation}
Here again, we are measuring all quantities in units of $r_H=r_0$. 
Using the expression for the charge diffusion constant in terms of conductivity 
given in (\ref{Dcval}), we find the conductivity for the D1-brane system in absence of 
charge density is given by 
\begin{equation}
\label{dcvalzero}
 \sigma = \frac{1}{16\pi G_3}.
\end{equation}

\vspace{.5cm}
\noindent
{\bf{ii. $q=0$ limit}}
\vspace{.5cm}

Note that the formula for conductivity as well as the Kubo's formula for 
bulk viscosity involves  the $q\rightarrow 0$ limit. It is therefore useful
to examine the fundamental equations in this limit.
The following simplifications occurs in this limit
\begin{equation}
 \alpha_t \rightarrow - \omega^2, \qquad
V \rightarrow -(3H+1) \omega^2.
\end{equation}
Examining  the equation  for the gauge field ( \ref{gpu}),  we see that it decouples from 
$Z_P$ and $S_P$ and it reduces to  
\begin{equation}
\label{Gpq0}
 G_P''+[K(4H-1)-H(2H+1)]\f{G_P'}{uHK}-\f{(H-1)(H-K)}{u^2H^2K}G_P+\f{L^6}{4}\f{H}{K^2}w^2G_P=0.
\end{equation}
The equation for $Z_P$ and $S_P$ are coupled, they reduce to 
\bea
\label{vissys}
Z_P''&+&\lf \{\f{K-2H-1}{u K}+\f{2(H-1)}{u H(3H+1)}\ri \}Z_P'+\f{L^6 \o ^2 H}{4K^2}Z_P\\ \nn
&=&\f{(H-1)(H-K)}{u^2HK(3H+1)}Z_P-\f{(9H^2-2H+1)-6(H-1)(H-K)}{3u^2H^2K(3H+1)}\tilde S_P, \\
\tilde S_P''&-&\f{\{K(H-2)+H(2H+1)\}}{uHK}\tilde S_P'+\f{6(H-1)}{u(3H+1)}Z_P'\nn\\
&=&\f{3(H-1)(H-K)}{u^2K(3H+1)}Z_P-\lf\{\f{L^6 \o^2H}{4K^2}+\f{2H}{u^2K}+\f{(H^2-1)(3H-2)}{u^2H^2(3H+1)}\ri\}\tilde S_P,  \nn
\eea
where 
\begin{equation}
 \tilde S_P = \omega^2 S_P.
\end{equation}
It is now possible to decouple the equations for $Z_P$ and $\tilde S_P$ 
by redefining $\tilde S_P$ as 
\begin{equation}
\label{newspd}
 \tilde S_P = \hat S_P + \frac{ 3H ( 1-H)}{3H+1} Z_P.
\end{equation}
In terms of $\hat S_P$, the equations in (\ref{vissys}) reduce to 
\bea
\label{decoupled}
\hat S_P''&-&\f{\{K(H-2)+H(2H+1)\}}{uHK}\hat  S_P'\nn\\
&=&-\lf\{\f{L^6\o^2H}{4K^2}+\f{2H}{u^2K}+\f{(H^2-1)(3H-2)}{u^2H^2(3H+1)}\ri\}
\hat S_P, \nn\\
Z_P''&+&\lf \{\f{K-2H-1}{u K}+\f{2(H-1)}{u H(3H+1)}\ri \}Z_P'-\f{(H-1)(H-K)}{u^2HK(3H+1)}Z_P +\f{L^6\o ^2 H}{4K^2}Z_P\nn\\ 
&=&-\f{(9H^2-2H+1)-6(H-1)(H-K)}{3u^2H^2K(3H+1)}
\lf(\hat S_P + \f{3H(1-H)}{3H+1} Z_P\ri ). 
\eea
Note that $Z_P$ decouples from the equation for $\hat S_P$.  We can now set 
$\hat S_P$ consistently to zero and study only the decoupled equation for 
$Z_P$. Simplifying the equation for $Z_P$, we obtain 
\bea
\label{eqnZ_P}
 Z_P''&+&\lf \{\f{K-2H-1}{u K}+\f{2(H-1)}{u H(3H+1)}\ri \}Z_P'+\f{L^6 \o ^2 H}{4K^2}Z_P\nn\\
&=&
 \frac{(H-1)( K(3H-7) + ( 3H+1) ( 2H+1))}{ u^2 HK(3H+1)^2}  Z_P.
\eea
Thus we have shown that setting $q=0$, we can obtain two decoupled 
equations (\ref{Gpq0}) and (\ref{eqnZ_P}) which correspond to the 
charge diffusion mode and the sound mode. 
Thus to obtain conductivity and the bulk viscosity of the 
charged D1-brane fluid, it is sufficient to study the 
equations (\ref{Gpq0}) and (\ref{eqnZ_P}). 

\vspace{.5cm}
\noindent
{\bf iii.  Behaviour at the horizon }
\vspace{.5cm}

To obtain the behaviour of the functions $G_P$ and $Z_P$ at the horizon, we 
define $x = \ln(1-u)$. Then both the  
equations  (\ref{Gpq0}) and (\ref{eqnZ_P}) 
reduce to the oscillator equation
\begin{equation}
 (\partial_x^2 + \frac{L^6}{4}\frac{1+k}{(2k+3)^2}  \omega^2 ) Y =0 .
\end{equation}
The ratio, $\frac{1+k}{(2k+3)^2}$ is obtained due to the behaviour of the 
coefficient proportional to $\omega^2$ in both the equations. 
Thus the  behaviour near the horizon is given by
\begin{equation}
 G_P, Z_P\rightarrow (1-u)^{\pm i \frac {L^3 \sqrt{1+k}}{ 2 (2k+3) }} , 
\quad \mbox{for} \quad u\rightarrow 1 .
\end{equation}
Since classically  horizons do not radiate, we need to choose the 
ingoing boundary condition 
\begin{equation}
\label{horbc}
(1-u)^{- i \frac {L^3 \sqrt{1+k}}{ 2 (2k+3) }},
\end{equation}
to solve these equations. 

\vspace{.5cm}
\noindent
{\bf iv. Behaviour at the boundary}
\vspace{.5cm}

Examining the coefficients of the equation for $G_P$ given in (\ref{Gpq0}) 
for $u\rightarrow 0$, the boundary, the equation reduces to 
\begin{equation}
 G_P'' + 2k G_P' + \frac{L^6}{4} \omega^2 G_P =0.
\end{equation}
Thus the solution for $G_P$ at the boundary, $u\rightarrow 0$, admits a Taylor series 
expansion of the form 
\begin{equation}
 G_P \sim A( 1 + O(u^2)   )  + B u ( 1+ O(u^2) )  +   \qquad u\rightarrow 0, 
\end{equation}
where $A$ and $B$ are integration constants. 
Similarly examining the coefficients of the equation for $Z_P$  given in (\ref{eqnZ_P}),
we see that, at the boundary, the equation reduces to 
\begin{equation}
  Z_P''   -\frac{2}{u} Z_p' + \frac{8k}{9u } Z_P =0.
 \end{equation}
 The above equation admits an expansion of the 
 form 
\begin{equation}
\label{zpexpbc}
 Z_P \sim  A ( 1+ \cdots  )   + Bu^3 ( 1+ \cdots ). 
\end{equation}
The behaviour at the boundary  is necessary to obtain the transport coefficients.
In fact, the transport coefficients are proportional to the 
ratio $B/A$, that is the ratio of the normalizable mode by the non-normalizable mode.

\section{Transport coefficients from gravity}

We first summarize the method put forward by 
\cite{Son:2002sd,Herzog:2002pc} to evaluate 
transport coefficients from gravity. 
\begin{enumerate}
\item Let $Z_k(r)$ be the gauge invariant variables constructed 
from the fluctuating gravity fields. 
 In general, they satisfy  coupled second 
order linear  differential equations. 
We choose  linear combination 
$Z(r)$  such that they   satisfy decoupled second order linear 
differential equations. These decoupled gauge invariant variables 
correspond to the hydrodynamic modes of the field theory. 
\item A local solution of the second order  differential equations 
near the horizon $r=r_0$ will in general be a superposition of 
incoming and outgoing waves. Classically the horizon
does not radiate, therefore, we choose the 
incoming wave boundary condition at the horizon.
\item The solution which obeys incoming wave boundary 
condition at the horizon can be 
written as a linear combination  of two local solutions
$f_1(r) $ and $f_2(r) $  at the boundary $r\rightarrow\infty$  as 
\begin{equation}
\label{boundsol}
Z(r) = A f_1(r) + B f_2(r),
\end{equation}
where $A$ and $B$  are the connection coefficients of 
the corresponding differential equations. 
Coefficients $A$ and $B$ depend on the parameters $\omega, q$ 
which enter the differential equation. Near the boundary, the 
solution (\ref{boundsol}) admits an expansion 
\begin{equation}
\label{boundexp}
Z(r) = A (1+\cdots)  + B r^{-\Delta} ( 1 + \cdots) ,
\end{equation}
where the ellipses denote higher powers of $r$ which are suppressed 
as $r\rightarrow\infty$ and $\Delta>0$. 
\item The action of the quadratic fluctuations can also be 
organized in terms of the gauge invariant observables. 
Evaluating the action on shell, it reduces to a boundary term
which is of the form 
\begin{equation}
S^{(2)} = \lim_{r\rightarrow\infty} \int d\omega dq 
F(r, \omega, q) Z'(r) Z(r) + {\mbox{contact terms}}, 
\end{equation}
where the contact terms do not involve derivatives of $Z(r)$ and
\begin{equation} 
\label{propcoef}
F(r, \omega, q)\rightarrow r^{\Delta +1} f(\omega, q), \qquad \mbox{as}, \qquad 
r\rightarrow \infty.
\end{equation} 
\item We can now use the fact that $Z(r)$ is a linear combination
of the fluctuation gravity fields and 
apply the prescription in \cite{Son:2002sd,Herzog:2002pc} to compute the retarded
correlator for the corresponding operator $O$ in the field theory. 
We obtain 
\begin{equation}
\label{derrel}
\langle OO\rangle_{R}  \sim \frac{B}{A} \sim
\left. \frac{r^{\Delta +1} }{Z} \frac{d Z(r)}{ dr}\right|_{r\rightarrow \infty, {\mbox{finite term}}}.
\end{equation}
We have not written an equality but used $\sim$
 as we have not yet kept track of the 
  proportionality constant which depends on $F(r, \omega, q)$ in the 
  limit $r\rightarrow\infty$.  We have used the expansion 
  in (\ref{boundexp}) and the property
  (\ref{propcoef})  to write the last relation in (\ref{derrel}).  
  Note that from the last expression in (\ref{derrel}), we need to extract the finite
  piece to obtain the ratio $B/A$. 
  \item
  To apply Kubo's formula, we only need the retarded correlator
  with $ q=0$. Thus it suffices to 
  evaluate the following  ratio to obtain the transport coefficients
  of interest, 
  \begin{equation}
  \lim_{r\rightarrow\infty}
  \left.  \frac{1 }{Z} \frac{d Z(r)}{ dr}\right|_{q=0}.
 \end{equation}
 In fact, for the DC conductivity and the bulk viscosity, we need to 
 take a $\omega \rightarrow 0$ limit which is given by 
 \begin{equation}
 \label{goodlimit}
 {\rm Re} \left( \lim_{r\rightarrow\infty, 
  \omega\rightarrow 0}
  \left.  \frac{1}{i \omega Z} \frac{d Z(r)}{ dr}\right|_{ q=0} \right). 
 \end{equation}
\end{enumerate}

Since it is only the ratio
\begin{equation}
\label{ratio}
{\cal R}(r)  = \frac{1}{Z} \frac{d Z(r)}{ dr}.
\end{equation}
at $r\rightarrow\infty$ that determines the retarded correlators, one can 
determine the differential equation satisfied by ${\cal R}(r)$ from the 
second order ordinary  linear differential equation satisfied by $Z(r)$. 
We will see that this is a first order,  ordinary 
but non-linear differential  equation. 
The boundary conditions for this differential equation
are determined from the ingoing boundary conditions satisfied by 
$Z(r)$ at the horizon. 
This equation, in fact, governs the radial evolution of the 
transport coefficients. 
We will show that for the DC conductivity and for the 
bulk viscosity, this equation is exactly solvable enabling us 
to determine the analytic expressions for these transport coefficients.
The fact that the evaluation of transport coefficients can be reduced to solving 
a first order but non-linear  differential equation has been observed recently 
for the case of ${\cal N}=4$ super-Yang Mills by \cite{Banerjee:2010zd}
and has been argued to be true in general in \cite{Bredberg:2010ky}. 

The rest of this section is organized as follows:
We first show that the radial evolution 
of the transport coefficients are determined by 
 first order non-linear ordinary differential equations. 
 These equations are exactly solvable for the DC conductivity 
 and the bulk viscosity. We then evaluate the effective action 
 to determine the proportionality constant relating the ratio
 ${\cal R}$ in (\ref{ratio}) to the transport coefficients.

\subsection{Radial evolution of the transport coefficients}

\vspace{.5cm}
\noindent
{\bf Radial evolution of conductivity}
\vspace{.5cm}

Let us obtain the equation that governs the radial evolution of
conductivity.
Note that  the equation for $G_P$ can be written as
\begin{equation}
\label{mingp}
\frac{1}{HK} ( HK G_P')' 
-\f{(H-1)(H-K)}{u^2H^2K}G_P+\f{L^6}{4}\f{H}{K^2}w^2G_P=0.
\end{equation}
One can now think of this mode as a minimally coupled scalar with a mass term 
proportional to $(H-1)$.   Thus except for the term proportional to $(H-1)$, 
it  falls in the class of equations of motion studied in \cite{Iqbal:2008by} for which 
the radial evolution of the transport coefficients was easy to obtain in the $\omega\rightarrow 0$ limit
\footnote{See equation (38) in \cite{Iqbal:2008by}.}.
To remove this term from (\ref{mingp}),  we perform the following redefinition
\begin{equation}
\label{newg}
 G_P = \frac{2H+1}{H} G.
\end{equation}
Then the equation for $G$ reduces to 
 \bea
\label{y1}
G''+\lf(\f{8H^2+1}{uH(2H+1)}-\f{2H+1}{uK}\ri )G'+\f{L^6}{4}\f{H}{K^2}w^2G=0.
\eea
Thus, the redefinition in (\ref{newg}) removes the mass term and reduces the equation 
to that of a minimally coupled massless scalar. 
To obtain the R-charge  retarded correlator, we need to impose ingoing boundary conditions 
at the horizon,  $u=1$ on $G_P$. From the redefinition in (\ref{newg}), we see that 
this translates to ingoing boundary condition on $G$. 
From the discussion in around (\ref{horbc}), we see that we have to impose  the condition
\begin{equation}
\label{horbc1}
G \sim (1-u)^{- i \frac {L^3 \sqrt{1+k}}{ 2 (2k+3) }}, \qquad u\rightarrow 1.
\end{equation}
As we have discussed earlier conductivity is proportional to the ratio
\begin{equation}
\label{cond1}
{\cal R}_{G_P} =  \frac{1}{i\omega G_P} \frac{d G_P}{du} = 
\frac{i}{\omega} \frac{k}{H (2H+1)} +  \frac{1}{i\omega G} \frac{d G}{du},
\end{equation}
where we have used the redefinition  given in (\ref{newg}) and also changed 
the variable from $r$ to $u$. 
Thus we need to evaluate the ratio $ \frac{1}{i\omega G} \frac{d G}{du}$
at the boundary subject to the condition (\ref{horbc1}) at the horizon. 
Let us define this ratio as 
\begin{equation}
f_{G} = \frac{1}{i\omega G} \frac{d G}{du}.
\end{equation}
The boundary condition for this ratio at the horizon, 
$u=1$  is then given by 
\be
\label{bcfr}
f_G|_{r_H}\ra\f{L^3\sqrt{1+k}}{2(3+2k)(1-u)}+\ldots,
\ee
where the ellipses refer to sub-leading terms at $u=1$. 
The differential equation satisfied by the ratio $f_G$ can be obtained from the
differential equation in (\ref{y1}).  This is given by
\begin{equation}
f_G' +\lf( \f{8H^2+1}{uH(2H+1)}-\f{2H+1}{uK}\ri )f_G
-i\frac{L^6 H}{4 K^2} \omega + i \omega f^2_G =0.
\end{equation}
This is a first order non-linear differential equation which governs the 
radial evolution of conductivity. 
From this equation, it is easy to obtain the DC conductivity and the pole
at $\omega\rightarrow 0$ present in the imaginary part of the conductivity. 
We first decompose  the above equation into its real and imaginary parts.
\bea
\label{feqn}
{\rm Re} f_G+\lf( \f{8H^2+1}{uH(2H+1)}-\f{2H+1}{uK}\ri )
{\rm Re} f_G  -2\o {\rm Im }f_G  {\rm Re} f_G &=&0, \nn\\
{\rm Im }f_G'+\lf(\f{8H^2+1}{uH(2H+1)}-\f{2H+1}{uK}\ri ){\rm Im f}_G-\o \lf(
{\rm Im} f_G^2- {\rm Re} f_G^2+\f{L^6H}{4 K^2}\ri )&=&0.\nn
\eea
These equations simplify and decouple in the limit $\o\ra 0$. 
This decoupling would not have been possible in the original equation for 
$G_P$ given in (\ref{mingp}) due to the presence of the mass term 
proportional to $(H-1)$.  But removing 
this term through the re-definition in (\ref{newg})
enables us to calculate DC conductivity exactly as follows.
The solution for $f_G$  satisfying the boundary condition (\ref{bcfr}) in
the $\omega\rightarrow 0$ limit is given by 
\bea
\label{solnf1}
{\rm Re}f_G=\f{(2k+3)^2}{\sqrt{1+k}} \f{H}{K (2H+1)^2 }, 
\qquad
{\rm Im} f_G = 0.
\eea
We can now use (\ref{cond1}) to evaluate the ratio which is 
proportional to the real part of the 
 DC conductivity. This is given by 
\bea
\label{dccondval}
{\rm Re } \left( {\cal R}_{G_P}\right)_{u\ra0, \o\ra0}
&=&\f{L^3}{2} \lf [{\tr Re}f_G+{\tr Re}\lf(\frac{i}{\omega} \frac{k}{ H(2H+1)} \ri)\ri ]_{u\ra 0,\o \ra 0}, \nn\\\label{condL3}
&=&\f{L^3}{2}\f{ (2k+3)^2}{9 \sqrt{1+k}}.
\eea
Now from the solution for ${\rm Im}  f_G$ given in (\ref{solnf1}) in the 
$\omega\rightarrow 0$ limit, we see that 
 ${\rm Im }f_G={\cal{O}}(\o)$.
Thus imaginary part of the conductivity in the $\omega\rightarrow 0$ 
limit is given by 
\be
\label{imcondval}
{\rm Im } \left( {\cal R}_{G_P}\right)_{u\ra0, \o\ra0} = 
{\rm Im}\lf (\frac{i}{\omega} \frac{k}{ H(2H+1)} \ri )_{u\ra 0}=\f{k}{3\o}.
\ee
Therefore we see that the imaginary part of the conductivity has a 
pole at $\omega\rightarrow 0$ limit which is 
expected because of  the translational invariance 
of the system. Translational invariance implies that there are no-impurities, 
which in turn implies infinite conductivity  at $\omega= 0$ by Drude's formula. In fact, using the Kramers-Kronig relation
\begin{equation}
{\rm Im}\, \sigma(\omega) = - \frac{1}{\pi} {\cal P} \int_{-\infty}^\infty
\frac{ {\rm Re}\, \sigma(\omega') }{\omega' - \omega } d\omega', 
\end{equation}
we see that the real part of the conductivity contains a delta function
if and only if the imaginary part has a pole. Since we have found a pole
in the imaginary part of the conductivity, it follows that the real part has 
a delta function singularity at $\omega =0$. Therefore the  value for the 
DC conductivity
\footnote{ Recently \cite{Jain:2010ip} has made a proposal for the value
of the DC conductivity for conformal systems with chemical potential in
arbitrary dimensions. We thank Sean Hartnoll for bringing
this reference to our attention.} 
is  valid at $\omega \rightarrow 0^+$. 

As a further check on our analytical manipulations, we have 
solved the differential equation for conductivity given in (\ref{mingp})
numerically
subject to the ingoing boundary conditions at the horizon and 
evaluated the ratio ${\cal R}_{G_P}$. 
For very small values of $\omega$, we find very good agreement 
with the formula given in (\ref{dccondval})  and 
(\ref{imcondval}). This is  shown in figure.  1  of 
section  5.2. 

\vspace{.5cm}
\noindent
{\bf Radial evolution of bulk viscosity}
\vspace{.5cm}

The bulk viscosity is determined from the equation for $Z_P$ given in 
(\ref{eqnZ_P}) which can be written as 
\begin{eqnarray}
\frac{1}{K} \frac{d}{dy} \left( K \frac{d Z_P}{dy} \right) 
+ \frac{2 (H-1) }{3y H( 3H+1) } \frac{dZ_P}{dy} 
+\f{ \o ^2 L^6H}{36 u^4K^2}Z_P\nn\\
=
 \frac{(H-1)( K(3H-7) + ( 3H+1) ( 2H+1))}{9  u^6 HK(3H+1)^2}  Z_P.
\end{eqnarray}
where $ y =u^3$. Again, we see that the equation 
resembles a minimally coupled scalar equation except for the 
terms proportional to $(H-1)$. We can remove these terms by the 
following redefinition for $Z_P$. 
\begin{equation}
\label{goodz}
 Z_P = \frac{3H+1}{H} Z.
\end{equation}
Then the equation for $Z$ reduces to the simple form
\begin{equation}
\label{goodeqz}
Z''+\f{K-2H-1}{uK}Z'+\f{\o ^2L^3H}{4K^2}Z=0.
\end{equation}
To obtain the retarded two point function of the stress tensor
we need to impose ingoing boundary condition on $Z_P$ at $u=1$. 
Using the redefinition in (\ref{goodz}), we see that
this translates to  the ingoing  boundary condition
on $G$  at the horizon. Therefore, we need to impose
\begin{equation}
Z \rightarrow (1-u)^{- i \frac {L^3 \sqrt{1+k}}{ 2 (2k+3) }}, \qquad u\rightarrow 1
\end{equation}
From  the earlier discussion, we see that the bulk viscosity is proportional to 
the real part of the following  ratio evaluated at the boundary. 
\begin{eqnarray}
{\rm Re} ( {\cal R}_{Z_P})  &=& { \rm Re} 
\left( \frac{1}{i\omega 3u^2 Z_P}\frac{ d Z_P}{du} \right),  \\ \nonumber 
&=& {\rm Re} \left(  \frac{1}{i\omega }\frac{k}{3u^2(1+ k u)} \right)  +
{\rm Re} \left( \frac{1}{i\omega 3u^2 Z}\frac{ d Z}{du} \right), \\ \nonumber
&=& {\rm Re} \left( \frac{1}{i\omega 3u^2}\frac{ d Z}{du} \right),
\end{eqnarray}
Here, we have used the redefinition of $Z_P$ given in (\ref{goodz}). 
 We are dividing by $3u^2$ so that we can extract out the 
ratio $B/A$ in the expansion of $Z_P$ near the boundary given in 
(\ref{zpexpbc}). Note that since  the bulk viscosity
is proportional to  
 the real part of the ratio $ \frac{1}{i\omega 3u^2}\frac{ d Z_P}{du}$, 
 it is  determined by the behaviour of 
$Z$.
Therefore let us define the ratio
\begin{equation}
 f_Z =  \frac{1}{i\omega Z } \frac{ dZ}{du}.
\end{equation}
Using the ingoing boundary condition for $Z$
at the horizon,  boundary condition for $f_Z$ at the horizon 
is given by
\begin{equation}
\label{bcbulk}
 f_Z|_{u\rightarrow 1} \rightarrow  \f{L^3\sqrt{1+k}}{2(3+2k)(1-u)}+\ldots, 
\end{equation}
where ellipses refer to sub-leading terms at $u=1$. 
The differential equation satisfied by $f_Z$ can  be  obtained from the differential equation 
for $Z$ in (\ref{goodeqz}) and is given by 
\begin{equation}
 f_Z' + \f{K-2H-1}{uK}f_Z  -i\frac{L^6 H}{4 K^2} \omega + i \omega f_Z^2 =0. 
\end{equation}
Again separating into the real and imaginary parts we obtain
\begin{eqnarray}
\label{realandimf}
{\rm Re}f_{Z}'+ \f{K-2H-1}{u K}{\rm Re} f_{Z} -2\o {\rm Im} f_{Z}{\rm Re}f_{Z}=0,\nn\\
{\rm Im }f_{Z}'+ \f{K-2H-1}{u K}{\rm Im}f_{Z} +
\o({\rm Re} f_{Z}^2-{\rm Im} f_{Z}^{2})-\f{\o L^6 H}{4K^2}=0.
\end{eqnarray}
The solution of these equations in the $\omega\rightarrow 0$ limit  which obeys the boundary
conditions in (\ref{bcbulk}) is given by 
\begin{equation}
 {\rm Re} f_Z =\frac{L^3\sqrt{1+k} u^2}{2 K} , \qquad {\rm Im} f_Z =0.
\end{equation}
It is now easy to obtain the ratio which is proportional to the bulk viscosity. It is given by
\begin{eqnarray}
\label{ratibulk}
 {\rm Re} ( {\cal R}_{Z_P}) |_{u\rightarrow 0} 
&=& \left. {\rm Re} \left( \frac{1}{i\omega 3u^2}\frac{ d Z}{du} \right)
\right|_{u\rightarrow 0},  \\ \nonumber
&=& {\rm Re} \frac{f_Z }{3u^2},\nn \\ \label{visL3}
&=& \f{L^3}{2}\f{\sqrt{1+k}}{3}.
\end{eqnarray}

Again as a further check on our manipulations, we evaluate the ratio
$ {\cal R}_{Z_P}$ directly by solving the differential equation
(\ref{eqnZ_P}) numerically subject to ingoing boundary conditions at the horizon. 
We find the result for $\omega\rightarrow 0$ in very good agreement
with the expression given in (\ref{ratibulk}). 
This is shown in figure 2 of section 5.2. 

Note that the problem of obtaining 
 the DC conductivity and the bulk viscosity has been reduced to solving
 a first order but non-linear differential equation. 
 This equation governs the radial evolution of the ratio
 which is proportional to the respective transport coefficient. 
In the $\omega\rightarrow 0$ limit, 
the solution  of these transport coefficients were easy to obtain exactly. 

\subsection{Comparison with numerical analysis}
In this section, we solve the equations of motion for the charge diffusion and sound mode numerically and find the transport coefficients. We will actually find the ratio ${\cal R}_{G_P}$ and ${\rm Re}\, {\cal R}_{Z_P}$ which 
is proportional to the conductivity and the viscosity. 
Further more, we work in a normalization in which $L^3 =2$ for convenience.
 Since we have analytic expressions for DC value of conductivity as well as viscosity at very small $\o$,  we can check our numerics with these results. 
We also know the exact expression for conductivity and viscosity in the limit $l=0$ and this gives us another check on our numerical results.
\FIGURE{
\begin{tabular}{cc}
\begin{minipage}{.4\textwidth}
\begin{center}
\includegraphics[width=.9\textwidth]{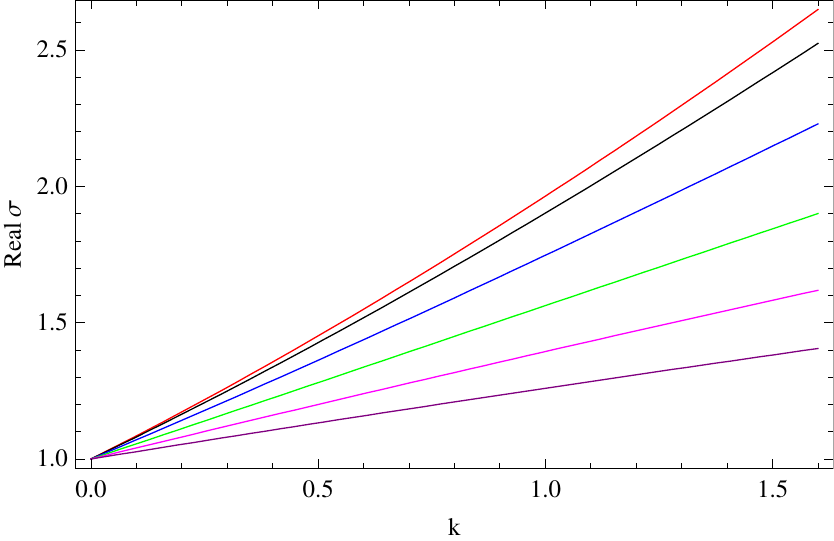}
\end{center}
\end{minipage}
\hspace{1 cm}
\begin{minipage}{.4\textwidth}
\begin{center}
\includegraphics[width=.9\textwidth]{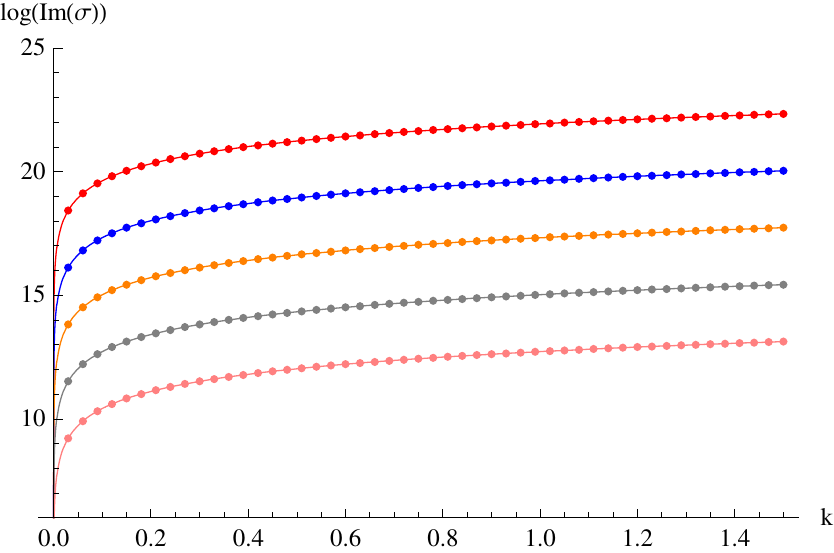}
\end{center}
\end{minipage}
\end{tabular}
\label{SCk}
\caption{Plots of real (on left) and log of imaginary part of conductivity vs k for the single charged case. On left, the different colors red, black, blue, green, magenta and purple correspond to $\o =$ $10^{-10}$, 0.2, 0.4, 0.6, 0.8 and 1.0 respectively. On right, the different colors red, blue, orange, gray and pink correspond to $\o =$ $10^{-10}$, $10^{-9}$, $10^{-8}$, $10^{-7}$ and $10^{-6}$ respectively.  The dots are the numerical values and the solid lines are curves Im $\s$ =  $\frac{k}{3\omega}$.  $\s$ is in units of $(16\pi G_3)^{-1}$ and $\o$ is in units of $2r_H^2/L^3$.}
}
\par
In figure (\ref{SCk}), we plot real and imaginary parts of conductivity  vs $k$  for the single charged case. For $k=0$, the real part of conductivity approaches 1. This is in accord with our analytic calculation for $l=0$ case. We compare the $k$ dependence obtained numerically for the real part of conductivity for very small $\o$ ($\o =10^{-10}$) with the DC conductivity. We find good agreement between them as the absolute value of the difference between numerical and analytically obtained values is at most $10^{-5}$.  We expect that the errors in our numerics remain in the same order for all other numerical curves, which tell dependence of AC conductivity on $k$ for different values of $\o$. We don't have analytic expressions for non-trivial $\o$ to compare them with. We find that conductivity increases monotonically with $k$, thought the slope decreases as we increase $\o$. Similarly we see that for small $\o$, our numerical results for imaginary part of conductivity fit well with the analytic expression. The absolute difference in this case is at most $10^{-6}$. For small $\o$, imaginary part of conductivity grows linearly with $k$.
\FIGURE{
\begin{tabular}{cc}
\begin{minipage}{.4\textwidth}
\begin{center}
\includegraphics[width=.9\textwidth]{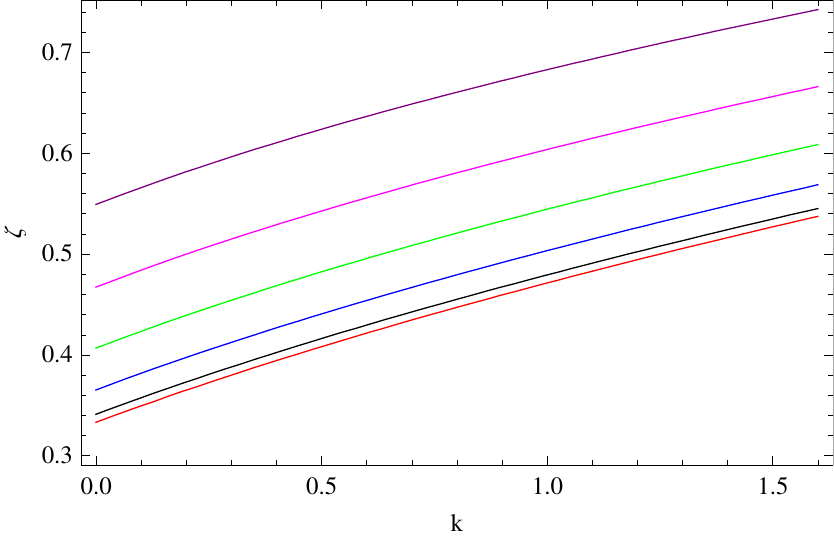}
\end{center}
\end{minipage}
\hspace{1 cm}
\begin{minipage}{.4\textwidth}
\begin{center}
\includegraphics[width=.9\textwidth]{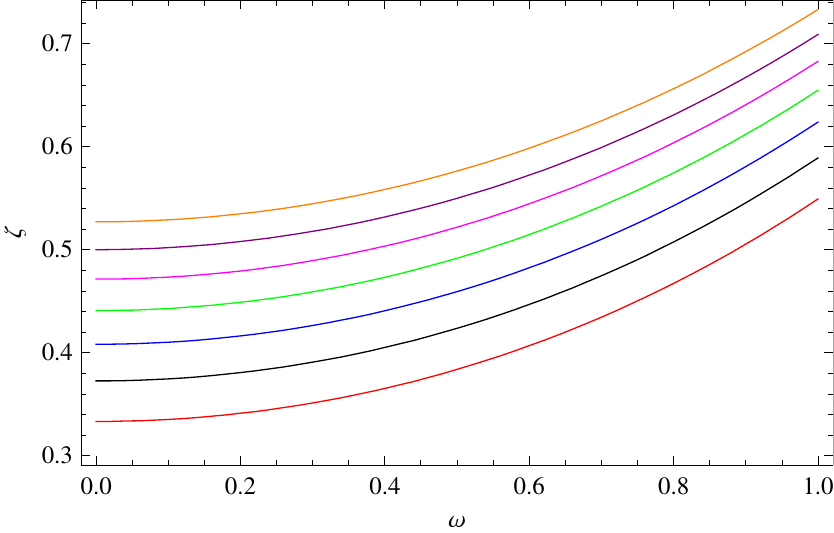}
\end{center}
\end{minipage}
\end{tabular}
\label{SCv}
\caption{Plots of viscosity vs k (on left) and $\o$ for single charged case. On left, the different colors red, black, blue, green, magenta and purple correspond to $\o =$ $10^{-10}$, 0.2, 0.4, 0.6, 0.8 and 1.0 respectively. On right,  the different colors red, black, blue, green, magenta, purple and orange correspond to $k=$ 0, 0.5, 1.0, 1.5, 2.0, 2.5 and 2.9 respectively. $\z$ is in units of $r_H^4/(16\pi G_3L^4)$ and $\o$ is in units of $2r_H^2/L^3$. }
}
\par
 In figure (\ref{SCv}), we plot viscosity against $k$ and $\o$ for the single charged case. We note from the plot that the smallest value of $\z$ is at $\o=k=0$. We also compare the analytic expression for the viscosity as a function of $k$ for $\o \ra 0$ with the numeric plot of $\z$ vs $k$ for $\o =10^{-10}$ (red curve in left plot in figure (\ref{SCv}) ). We find the absolute difference between analytical and numerical values to be less than $10^{-4}$. From the curves, we see that the curve for $\z$ vs $k$ for a given $\o$ shifts as a whole as one changes $\o$. From the right plot, we see that the amount of shift increases non-linearly with $\o$.   
\FIGURE{
\begin{tabular}{cc}
\begin{minipage}{.4\textwidth}
\begin{center}
\includegraphics[width=.9\textwidth]{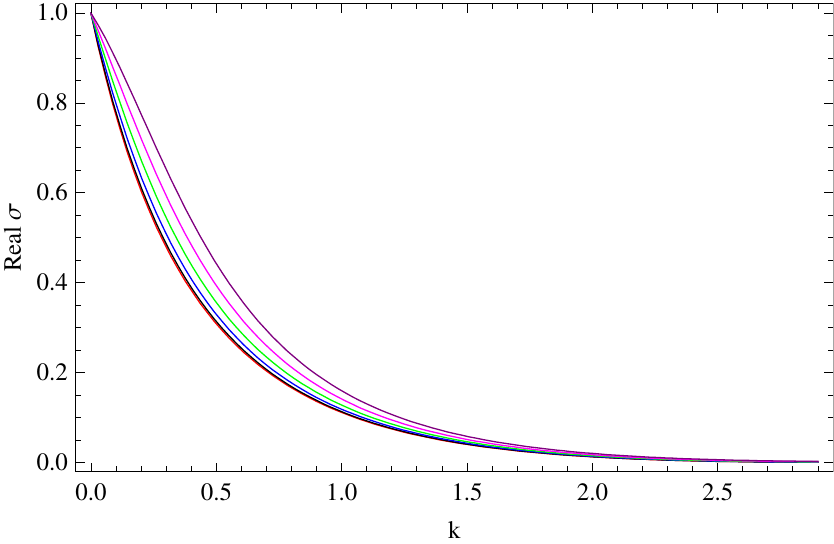}
\end{center}
\end{minipage}
\hspace{1 cm}
\begin{minipage}{.4\textwidth}
\begin{center}
\includegraphics[width=.9\textwidth]{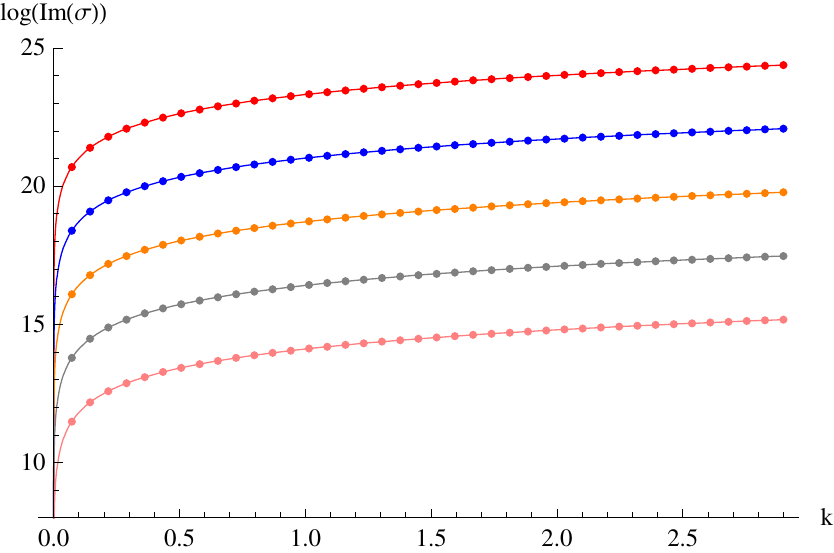}
\end{center}
\end{minipage}
\end{tabular}
\label{ECc}
\caption{Plots of real (on left) and log of imaginary part of conductivity vs k for the equal charged case. On left, the different colors red, black, blue, green, magenta and purple correspond to $\o =$ $10^{-10}$, 0.2, 0.4, 0.6, 0.8 and 1.0 respectively. On right, the different colors red, blue, orange, gray and pink correspond to $\o =$ $10^{-10}$, $10^{-9}$, $10^{-8}$, $10^{-7}$ and $10^{-6}$ respectively.  The dots are the numerical values and the solid lines are curves Im $\s$ =  $\frac{4k}{3\omega}$. $\s$ is in units of $(16\pi G_3)^{-1}$ and $\o$ is in units of $2r_H^2/L^3$. }
}
\par
In figure (\ref{ECc}), we plot the real and imaginary part of conductivity against $k$ for the equal charged case. Here too, we have analytic expressions for the DC conductivity which we compare with the  $\s$ vs $k$ plot for $\o=10^{-10}$, red curve in the left plot. We find a good agreement with the absolute difference between the numeric and analytic values being less than $10^{-6}$. This bound on error is also same for the plots of imaginary conductivity vs $k$ on the right. We observe here that there is little change in the curves of $\s$ vs $k$ as one changes $\o$. The $\s$ vs $k$ behaviour here is very different from the same in single charged case. Latter, the curves were monotonically increasing, but here, conductivity decreases with increasing $k$. 
\FIGURE{
\begin{tabular}{cc}
\begin{minipage}{.4\textwidth}
\begin{center}
\includegraphics[width=.9\textwidth]{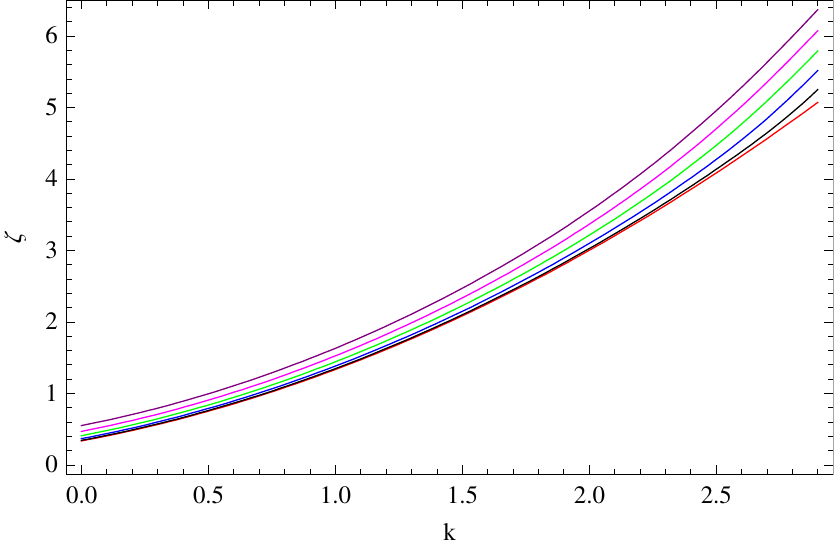}
\end{center}
\end{minipage}
\hspace{1 cm}
\begin{minipage}{.4\textwidth}
\begin{center}
\includegraphics[width=.9\textwidth]{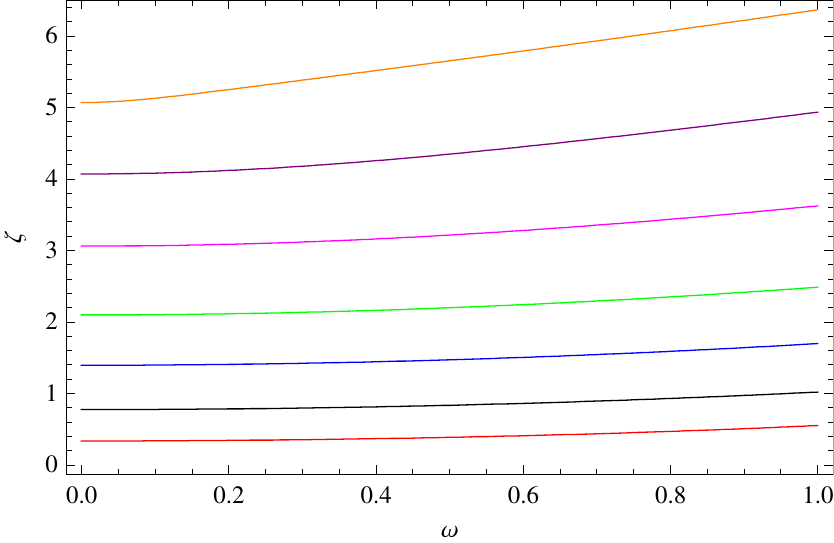}
\end{center}
\end{minipage}
\end{tabular}
\label{Eqv}
\caption{Plots of viscosity vs k (on left) and $\o$ for equal charged case. On left, the different colors red, black, blue, green, magenta and purple correspond to $\o =$ $10^{-10}$, 0.2, 0.4, 0.6, 0.8 and 1.0 respectively. On right,  the same correspond to $k=$ 0, 0.5, 1.0, 1.5, 2.0 and 2.5 respectively. Orange curve on right correspond to $k=2.9$. $\z$ is in units of $r_H^4/(16\pi G_3L^4)$ and $\o$ is in units of $2r_H^2/L^3$.}
}
\par
Now we plot the behaviour of viscosity vs $k$ and $\o$ in figure (\ref{Eqv}). Again as in the single charged case, the minimum value of viscosity is at $k=\o =0$, which is same as before and saturates the conjectured lower bound on bulk viscosity. The red curve on the left, which stands for viscosity vs $k$ at $\o =10^{-10}$ is compared to analytic value of viscosity obtained in the  $\o\ra 0$ limit. We get the absolute difference between analytic and numerical values to be less than $10^{-9}$ here, suggesting excellent agreement. We find little dependence of  viscosity on $\o$, particularly at smaller values of $k$.



\subsection{Evaluation of the transport coefficients} 

In this section, we use the standard prescription of the gauge/gravity 
correspondence to evaluate the retarded two point functions which 
determines the conductivity and bulk viscosity. 
This will determine the proportionality constant between the ratios
${\cal R}_{G_P}$, ${\cal R}_{Z_P}$ and the transport coefficients. 
For this, we first need to expand the bulk action given in 
(\ref{truncact}) along with the Gibbons-Hawking boundary term to second order 
in the fluctuations $H_{\mu\nu}, B_\mu, \varphi$ and $\xi$. 
In this section, we will not be using the dimensionless variables
given in (\ref{dimension}). All quantities in this section will have their 
respective dimensions, whenever needed, we will restore the
dimensions of the ratios ${\cal R}_{G_P}$ and ${\cal R}_{Z_P}$. 
Using equations of motion and the constraints (Equations (\ref{EqHzz}) to (\ref{New3})), we can write the bulk action (\ref{truncact}) 
expanded to second order in fluctuations as a total derivative in $r$.
\bea
S_{\tr{bulk}}^{(2)}&=&\f{1}{16\pi G_3}\int d\o dq dr\f{d{\cal{L}}_{B}}{dr}, \nn\\
\f{L^7}{r^7}{\cal{L}}_{B}&=&
\f{K}{4} \left(2
   {H_{tt}} {H_{tt}}'-{H_{zz}}
   {H_{tt}}'-{H_{tt}} {H_{zz}}'+2 {H_{zz}}
   {H_{zz}}'\right)
 -\frac{3}{2}H{H_{tz}}
    H_{tz}'\nn\\&&
-\frac{r}{2\o}(H-1)(H-K)
   \left(q{B_t}+w {B_z}\right)
   \left(r^3 H K B_z'+2H_{tz}\right)-\frac{8}{9} K \vp\vp'\nn\\&&
 -\frac{K}{2H^2}\xi\xi '+\frac{K}{3H}(\xi\vp'+\vp\xi ')+\frac{K}{r H^3}(2-H)\xi ^2 +\frac{2K}{3
   rH^2} (H-1)  {\vp} \xi  
\nn\\&&
 -\frac{1}{2r H}H_{tt} H_{zz} \{H (2
   H+1)+(4 H+1) K\} \nn\\&&
 +\frac{1}{4rH}
   ({H_{tt}}+{H_{zz}})  \left\{K(3 H+1)
   {H_{zz}}+H (2
   H+K+1)H_{tt}\right\}
\nn\\&&
+\frac{1}{rK}{H_{tz}}^2
   \{H (2 H+1)-(5 H+2) K\}
-\frac{4K}{3r}
   ({H_{tt}}-{H_{zz}}) {\vp}
\nn\\&&
  +\frac{ K}{3r
   H}\{(7 H+1) {H_{tt}}-(H-1) {H_{zz}}\} {\vp}
  -\frac{ K}{2
  r H^2}({H_{tt}}+{H_{zz}}) \xi.
\eea  
Note that here
the prime 
 denotes derivative with respect to $r$. 
 The Gibbons-Hawking term expanded to second order in fluctuations  is given by
\bea
S_{{\rm GH} }^{(2)}&=&\f{1}{8\pi G_3}\int d^2x \sqrt{-h}K_{ext},\nn\\
\f{8L^7}{r^6}\sqrt{-h}K_{ext}&=&
\f{4}{K}\{K(8H+3)-H(2H+1)\}H_{tz}^2 + 
 8 r H H_{tz} H_{tz}'\nn\\&&
-\f{1}{H}(H_{tt} - H_{zz})^2 \{K (1 + 4 H) + H (2 H + 1)\}\nn\\&&
 - 2 r K(H_{tt}- H_{zz}) (H_{tt}' - H_{zz}').
\eea
We  now combine the $S^{(2)}_{\rm{bulk}}$ and $S^{(2)}_{{\rm GH}}$.
Using the constraints, we can rewrite it in terms of the 
gauge invariant quantities $Z_P, G_P$ and  $S_P$  as follows
\bea
S&=&S^{(2)}_{\tr{bulk}}+S^{(2)}_{GH}=\f{1}{16\pi G_3}\int d\o dq {\cal {L}}, \nn\\
\f{L^7}{r^7}{\cal{L}}&=&\f{l^2r_0^6K}{2r^8\a_t H}\lf (\f{q}{V}Z_P+r^2HG_P\ri )\lf (\f{q}{V}Z_P+r^2HG_P\ri )'-\f{3H^2K}{2V^2}Z_PZ_P'\nn\\&&
-\f{K}{2}\lf (\f{Z_P}{V}+\f{S_P}{3H}\ri )\lf (\f{Z_P}{V}+\f{S_P}{3H}\ri )'+ {\tr{ contact terms}}.
\eea
where  `contact terms' represent those terms in the action which do not contain any derivatives in $r$ and the counter terms which render the complete 
boundary action finite. 
Next we define a new variable
\be
\cS =S_P+\f{3H(1-H)}{V}Z_P.
\ee
Note that this is also a gauge invariant variable. 
It has the following useful property
\begin{equation}
{\cal S} \rightarrow \frac{\hat S}{\omega^2}, \qquad \mbox{as} \quad q\rightarrow 0.
\end{equation}
where $\hat S$ is defined in (\ref{newspd}). 
Thus on taking $q\rightarrow 0$ limit, we can consistently set 
$\cal S$ to zero. 
 We can now rewrite the boundary Lagrangian using ${\cal S}$ as 
\bea
\f{L^7}{r^7}{\cal{L}}&=&\f{l^2r_0^6HK}{2r^4\a_t }\lf (\f{q}{r^2VH}Z_P+G_P\ri )\lf (\f{q}{r^2VH}Z_P+G_P\ri )'-\f{2H^2K}{V^2}Z_PZ_P'\nn\\&&
-\f{K}{18H^2}\cS\cS '-\f{K}{6V}(Z_P\cS'+\cS Z_P')+ {\tr{ contact terms.}}
\eea 
To evaluate the transport coefficients using the Kubo's formula in 
(\ref{kubo1}) and (\ref{kubo2}),
 it is sufficient to look at the boundary Lagrangian at $q\rightarrow 0$ limit. 
In this limit,
 we can set consistently $\cS=0$. So the boundary  Lagrangian can now be simplified 
 as 
 \bea
\label{beaq0}
{\cal{L}} =-\f{r^7}{L^7} \lf\{\f{l^2r_0^6}{2r^4\o ^2}G_{P}G_{P}'+\f{1}{8\o ^4}Z_PZ_P'+{\tr{contact terms}}\ri \}.
\eea
At $q=0$, the expression for $G_{P}$ reduces to 
\be
G_{P}=\o B_{z}=\o \f{L^2}{lr_{0}^3}A_{z}.
\ee
Substituting this in (\ref{beaq0}), the boundary action involving the gauge field can be
written as
\begin{equation}
\label{bcactg}
S^{(2)}_{{\rm gauge}} = \frac{r_H^2}{16\pi G_3L^3} \int d\omega dq (A_z^{(0)})^2 
i\omega{\cal R}_{G_P}|_{u=0}. 
\end{equation}
Here we have converted the derivative in $r$ to derivative in $u$
and used the definition of ${\cal R}_{G_P}$.  $A_z^{(0)}$ refers to the boundary 
value of the gauge field. 
This field couples to the R-current of the D1-brane theory by the coupling
\begin{equation}
S_{{\rm coupling}} = i \int d^2x  ( J^t A_t^{(0)} + J^zA_z^{(0)} ).
\end{equation}
Then using the gauge/gravity prescription, we can obtain the 
retarded Green's function of the R-current by
\begin{equation}
G_{zz} =  -\f{\delta^{2}S^{(2)}}{\delta A^{0}_{z}\delta A^{0}_{z}}.
\end{equation}
Using this prescription  and the boundary
action for the gauge field given in 
(\ref{bcactg} ), we obtain the following expression for the R-current
correlator from gravity
\begin{equation}
G_{zz} = -\frac{2 r_H^2}{16\pi G_3 L^3} i\omega {\cal R}_{G_P}|_{u=0}.
\end{equation}
Finally  we can compute the DC conductivity using the Kubo's formula
\bea
\sigma_{DC}&=& {\rm Re} \left( \lim_{\o\rightarrow 0}\f{i}{\o}G_{zz}(\o,q=0)
\right) ,\nn\\
&=&  \frac{2 r_H^2}{16\pi G_3 L^3}
\lim_{\o\rightarrow 0}{\rm Re} {\cal R}_{G_P}|_{u=0}, \nn\\
&=&\f{2r_{H}^2}{16\pi G_{3}L^3}\lf (\f{L^3}{2r_H^2}\f{(2k+3)^2}{9\sqrt{1+k}} \ri ), \nn\\
&=& \f{1}{16\pi G_{3}}\f{(2k+3)^2}{9\sqrt{1+k}}.
\eea
Here,  in the third step, we have used the result (\ref{dccondval}) 
and reinstated the proper dimensions for the ratio ${\cal R}_{G_P}$ which 
has the dimensions of length.  As a check of the final answer note that 
at $k=0$, 
it reduces to the value evaluated using the quasi-normal mode analysis
in (\ref{dcvalzero}).

Similarly we can determine viscosity using Kubo's formula.  
At $q=0$, the 
fluctuation $Z_P$ reduces to 
\begin{equation}
Z_P = \omega^2  H_{zz}  +2\frac{ (3H+1) }{3H} \varphi.
\end{equation}
Substituting this in (\ref{beaq0}), 
the boundary action involving quadratic terms in the fluctuation $H_{zz}$ is given by
\begin{equation}
\label{quadh}
S^{(2)}_{H_{zz}} = \frac{1}{16\pi G_3} \frac{3r_H^6}{4L^7}
\int d\omega dq ( H_{zz}^{(0)})^2 i\omega 
{\cal R}_{Z_P}|_{u=0}.  
\end{equation}
Again we have converted the derivative in $r$ to a derivative in $u$ and used 
the definition of ${\cal R}_{Z_P}$.  $H_{zz}^{(0)}$ refers to the boundary value of the 
fluctuation.  The boundary 
fluctuations of the metric  couples with the stress tensor of the field theory  by the following 
action \cite{Policastro:2002tn}
\begin{equation}
 S_{\rm coupling} = \frac{i}{2} \int d^2x ( H_{tt}^{(0)} T^{tt} + H_{zz}^{(0)} T^{zz} 
+ 2 H_{tz}^{(0)} T^{tz}). 
\end{equation}
Then using the standard gauge/gravity prescription, the two point function of the 
stress tensor is given by 
\be
\label{stressco}
G_{zzzz}=-4\f{\delta^{2}S^{(2)}}{\delta H_{zz}(\o)\delta H_{zz}(-\o)}.
\ee
Using this prescription and the quadratic action 
for the metric fluctuation given in (\ref{quadh}),  we see the above two point function
is given by 
\begin{equation}
 G_{zzzz} = - \frac{1}{16\pi G_3} \frac{3 r_H^6}{L^7} i\omega 
{\cal R}_{Z_P}|_{u=0}.  
\end{equation}
We now can compute the bulk viscosity using the Kubo's formula
\bea
\zeta&=& {\rm Re} \lf(\lim_{\o\rightarrow0}\f{i}{\o}G_{zz,zz}(\o,q=0)\ri), \nn\\
&=&  \frac{1}{16\pi G_3} \frac{6 r_H^6}{L^7} \lim_{\o\rightarrow0}
{\rm Re}\,   {\cal R}_{Z_P} |_{u=0}, \nn\\
&=& \frac{ r_H^4}{16\pi G_3 L^4} \sqrt{1+k}, \nn \\
&=& \frac{1}{4\pi} s. 
\eea
Here again, in the third line, we have used the expression for ${\cal R}_{Z_P}$  given in 
(\ref{ratibulk}). In the last line, we have written the expression for $\zeta$ using the definition 
of entropy density for the single charged D1-brane given in (\ref{therm1}). 
Thus we see that the ratio of bulk viscosity to entropy density 
remains $1/4\pi$ when  the charge density is turned on.

\section{Properties of the transport coefficients}

We first summarize the results of the transport coefficients of the single charged D1-brane.
\begin{eqnarray}
\label{valtransp}
 \sigma = \f{1}{16\pi G_3}\f{(2k+3)^2}{9 \sqrt{1+k}}, \\ \nonumber
\zeta = \frac{ r_H^4}{16\pi G_3 L^4} \sqrt{1+k}.
\end{eqnarray}
In this section, we restrict ourselves to only the DC conductivity except in subsection (\ref{condsection}). 
Using these two results, we can find three more transport coefficients. 
The charge diffusion constant is related to conductivity by (\ref{Dcval}) and is given by 
\begin{equation}
\label{chrgdiffval}
 D_c = \frac{L^3}{r_H^2} \frac{3-2k}{6 \sqrt{1+k}}.
\end{equation}
The thermal conductivity is also related to the conductivity by (\ref{thermcond}) 
and is given by 
\begin{equation}
\label{tc-1}
 \k _T=\lf (\f{\varepsilon +p}{\rho}\ri )^2\f{\s}{T}=\f{r_H^2}{8LG_3}\f{(2k+3)(1+k)}{k}.
\end{equation}
Finally the sound diffusion constant can be obtained by (\ref{soundspeed}) 
and is given by
\begin{equation}
D_s = \frac{\zeta}{2(\epsilon +p)} = \frac{L^3}{12 r_H^2 \sqrt{1+k}}. 
\end{equation}

As we have noted earlier, the ratio of bulk viscosity to entropy density is independent of 
the chemical potential and is given by
\begin{equation}
 \frac{\zeta}{s} = \frac{1}{4\pi}. 
\end{equation}
This property also holds  for the equal charged D1-brane solution as shown in appendix B. 
Using the formula for the bulk viscosity (\ref{valtransp}),  the thermal conductivity 
in (\ref{tc-1}),  the Hawking temperature  in (\ref{therm1})  and the chemical  potential in (\ref{therm3}), 
 we can show the following 
relationship between these quantities is true
\begin{equation}
\label{wfrel}
 \f{\k _T\mu ^2}{\z T}=\lf (2\pi L\ri )^2.
\end{equation}
This relationship is more striking when we write the chemical potential $\mu$ in terms of 
its dimensions. Note that the normalization of the gauge field we have used in 
(\ref{truncact}) is such that it is dimensionless. This is convenient for the gravity analysis, but 
it is conventional for the gauge field to have dimensions of inverse length. 
Since the chemical potential is basically the value of the gauge field at the horizon (\ref{therm3}), 
it must have the dimensions of inverse length. 
Let us therefore restore its dimensions by defining
\begin{equation}
 \hat \mu = \frac{\mu}{L}. 
\end{equation}
Then the relationship in (\ref{wfrel}) can be written as
\begin{equation}
  \f{\k _T\hat\mu^2}{\z T}= 4\pi^2. 
\end{equation}
This relationship is similar to the Wiedemann-Franz law seen between 
thermal conductivity and electrical conductivity. A similar relationship between 
thermal conductivity and the shear viscosity for the single charged D3 brane was observed 
by \cite{Son:2006em}.

\subsection{Transport coefficients at criticality}

In this section, we discuss the reason for this property as well as  behaviour of the 
transport coefficients  near the boundary of thermodynamic stability $k=3/2$. 
We first note that the charge diffusion constant $D_c$ for the 
single charged D1-brane given in (\ref{chrgdiffval}) vanishes at the boundary of 
thermodynamic instability. This indicates that this mode becomes unstable 
at $k=3/2$ and for this case the thermodynamic instability can be studied 
by examining this mode more carefully. 
As we will see in appendix B, this feature does not hold for the equal charged 
D1-brane. It was also not seen in the analysis of \cite{Son:2006em} for the single charged D3-brane. 
Thus this  feature seems to be specific for the single charged D1-brane and it is worth 
exploring this further. 

To determine the critical behaviour of the transport coefficients at the boundary of thermodynamic 
instability, we follow the analysis done by \cite{Son:2006em}. 
We first define the dimensionless chemical potential $\cm$ as
\be{\label{chemm}}
\cm= \frac{\hat\mu}{2\pi T_H} = \f{\mu}{2\pi L  T_H}=\f{\sqrt{k}}{(3+2k)}.
\ee
Note that $\mu/T$ is the natural variable that occurs in charge current (\ref{fluidstress}).  
We can invert the relation in (\ref{chemm})   to write $k$ as
\be
k=\f{1-12\cm^2-\sqrt{(1-24\cm^2)}}{8\cm^2}.
\ee 
Thus, we can re-express the transport coefficients as
\begin{eqnarray}
\label{scalform}
\sigma &=& \frac{1}{16\pi G_3 } 
\lf (\frac{1 - 12\cm^2 - \sqrt{ 1- 24\cm^2} }{ 72\sqrt{2} \cm^4} \ri )\left( \frac{
1- 4\cm^2 + \sqrt{ 1- 24\cm^2}}{ 1+ \cm^2} \right)^{1/2}, \\ \nonumber
\zeta &=& \frac{\pi L^2 T^2 }{4 G_3} \lf ( \frac{  1+ 6 \cm^2 + \sqrt{ 1- 24\cm^2} }{18}\ri )
\left(\frac{ 1- 4\cm^2 - \sqrt{ 1- 24\cm^2}}{ 8\cm^2} \right)^{1/2}, \\ \nonumber
\kappa_T &=&  \frac{\pi L^2 T }{4 G_3} \lf ( \frac{  1+ 6 \cm^2 + \sqrt{ 1- 24\cm^2} }{18\cm ^2}\ri )
\left(\frac{ 1- 4\cm^2 - \sqrt{ 1- 24\cm^2}}{ 8\cm^2} \right)^{1/2} , \\ \nonumber
D_c &=& \frac{1}{24\pi T} \sqrt{ 1- 24\cm^2} 
\left( \frac{  1+ 6 \cm^2 - \sqrt{ 1- 24\cm^2} }{\cm^2( 1+ \cm^2)}\right), \\ \nonumber
D_s &=& \frac{1}{48\pi T}  \left( \frac{5 + \sqrt{ 1- 24\cm^2} }{1+ \cm^2} \right). 
\end{eqnarray}
The  boundary of thermodynamic stability 
 lies at $k=\f{3}{2}$ or $\cm_c=\f{1}{\sqrt{24}}$. Expanding the transport coefficients near this point, we see that the $D_c$ exhibits a square root branch cut 
 at the critical point. 
 The other transport coefficients are finite at the critical point $\cm_c$, however
 their first derivatives including that of $D_c$ diverges as $(\cm_c - \cm )^{-1/2}$. 
 Thus the  critical index is $1/2$  which indicates that the system exhibits  mean field
 behaviour.  
A similar behaviour was observed for the shear viscosity and 
conductivity for the single charged D3 branes in \cite{Son:2006em}.

From the above expressions
for the transport coefficients in (\ref{scalform}), note that  that $\zeta$ and $\kappa_T$ 
are written as $T^2 f(\cm)$ and $Tg(\cm)$ respectively. 
This  demonstrates that the system has a hidden
$2+1$ conformal invariance since the entropy density is proportional to $T^2$ in 
$2+1$ dimensions.  Also note that the charge and sound diffusivity can be written 
in the scaling form $\frac{1}{T} f(\cm)$. The conductivity just depends on the 
dimensionless ratio $\cm$ and assumes the scaling form $f(\cm)$. 
From examining the scaling form, it is easy to see that as $T\rightarrow \infty$, 
keeping the chemical potential $\mu$ constant, all the expressions for the 
transport coefficients reduce to the uncharged case as expected. 
Another point worth mentioning is that  on expressing 
$G_3, L$ in terms of the Yang-Mills coupling and the rank
$N$, the transport coefficients
$\zeta$ and $ \kappa_T$ are proportional to $N^2/\sqrt{\lambda}$. 
If at all this system holographically describes a 
$1+1$ dimensional system seen in nature, 
the scaling behaviour of the transport coefficients
seen in  (\ref{scalform}) is a possible test. 

\subsection{Behaviour of conductivity} \label{condsection}

\FIGURE{ 
\begin{tabular}{cc}
\begin{minipage}{.4\textwidth}
\begin{center}
\includegraphics[width=.9\textwidth]{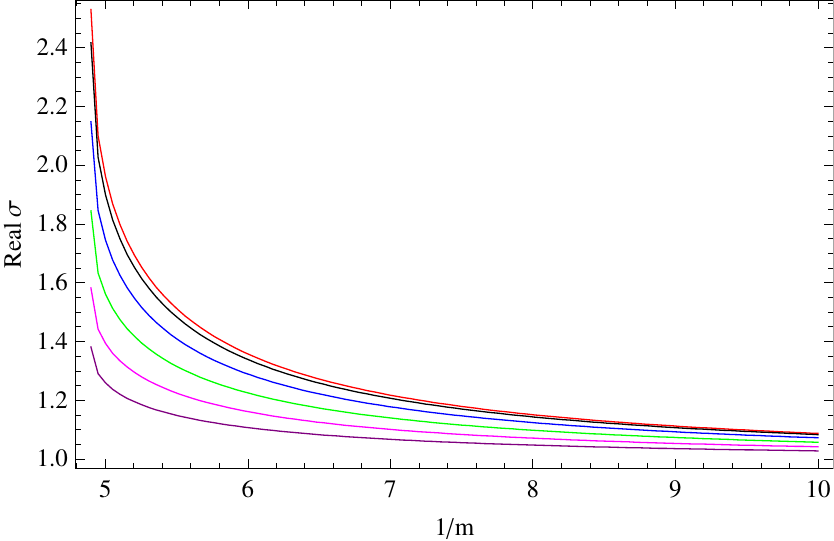}
\label{lnsvk}
\end{center}
\end{minipage}
\hspace{1 cm}
\begin{minipage}{.4\textwidth}
\begin{center}
\includegraphics[width=.9\textwidth]{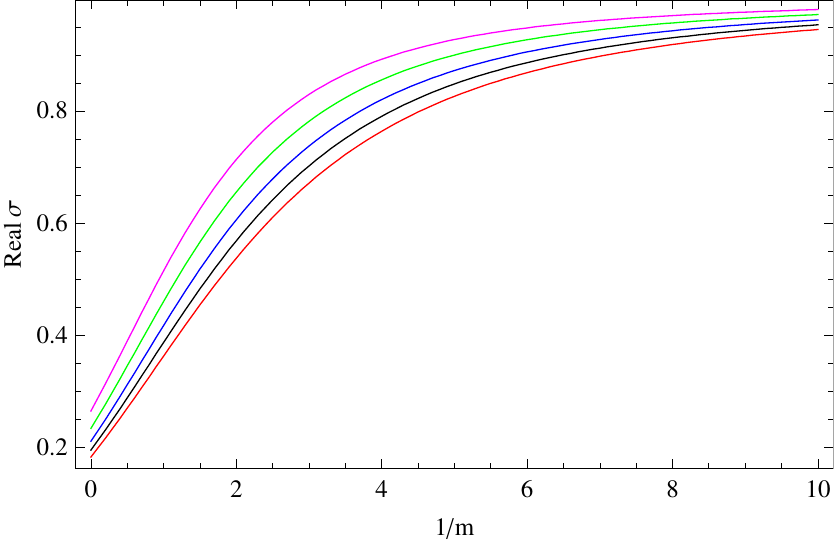}
\label{2Drsvw}
\end{center}
\end{minipage}
\end{tabular}
\label{Cvcm}
\caption{Plots of real part of conductivity vs $1/\cm$ for single charged (on left) and equal charged case. The different colors in the left plot, red, black, blue, green, magenta and purple
correspond to $\omega$ = $10^{-10}$, 0.2,  0.4 , 0.6, 0.8 and 1
respectively. The colours in the right plot, red, black, blue, green and magenta
correspond to $\omega$ = $10^{-10}$, 0.4 , 0.6, 0.8 and 1 respectively. $\s$ is in units of $(16\pi G_3)^{-1}$.  }
}
 In figure (\ref{Cvcm}), we plot the conductivity vs quantity $1/{\cm}$, which is proportional to temperature, if chemical potential is held constant. For the single charged case, we can't go to lower values of $1/\cm<1/\cm _c$. We note that for both the single charged case and equal charged case, the conductivity saturates to $1$, as temperature increases. This is expected from our uncharged brane analysis. As $\cm \ra 0$, the behaviour of DC conductivity is 
\bea
16\pi G_3\s_{DC}&\ra &1+\f{15}{2}\cm ^2+\cdots\quad {\tr{for single charged case}},\nn\\
&& 1-24\cm^2+\cdots \quad{\tr{for equal charged case.}}
\eea 
The low temperature behaviour in equal charged case is $\s_{DC}\sim \cm ^{-2}\sim T^2$ for $\cm^{-1} \ra 0$.   
\FIGURE{
\begin{tabular}{cc}
\begin{minipage}{.4\textwidth}
\begin{center}
\includegraphics[width=.9\textwidth]{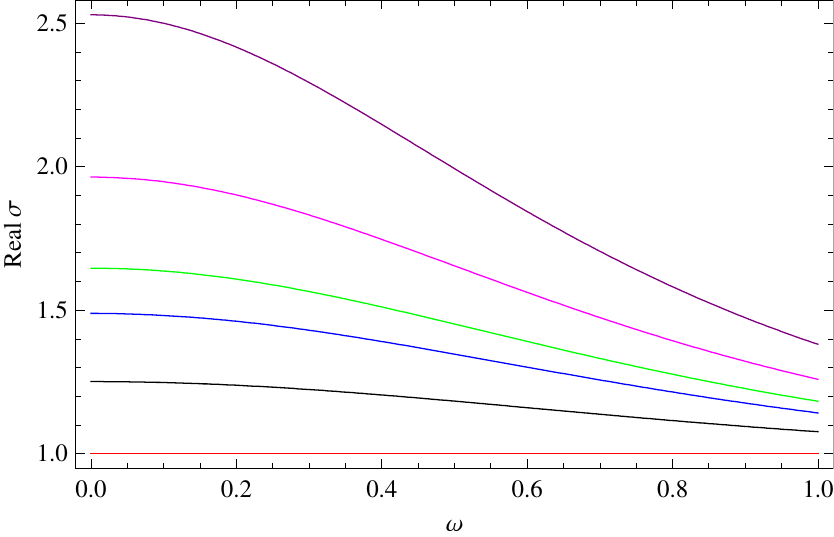}
\label{lnsvk}
\end{center}
\end{minipage}
\hspace{1 cm}
\begin{minipage}{.4\textwidth}
\begin{center}
\includegraphics[width=.9\textwidth]{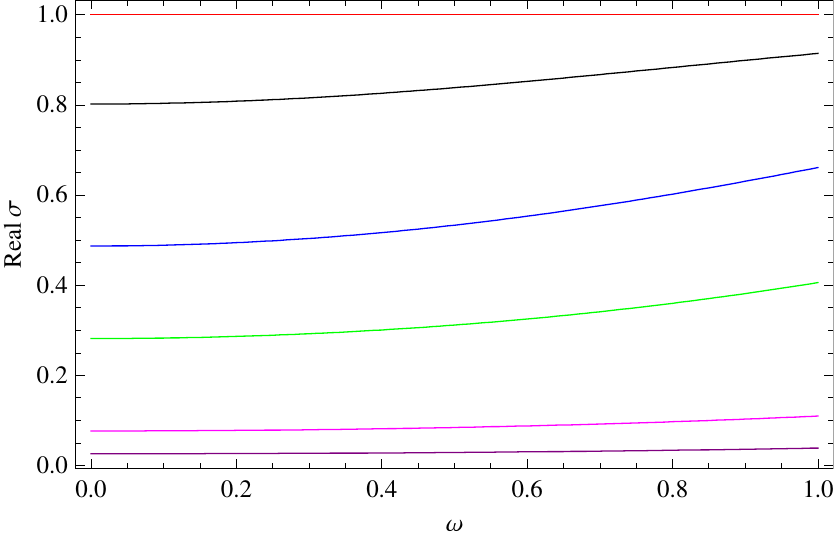}
\label{2Drsvw}
\end{center}
\end{minipage}
\end{tabular}
\label{Cvw}
\caption{Plots of real part of conductivity vs $\o$ for the single charged (on left) and equal charged case. The different colours in the left plot, red, black, blue, green, magenta and
purple correspond to $\cm$ = $10^{-3}$, 0.15, 0.18, 0.19, 0.2, 2$\sqrt{6}$
respectively. The different colours in the right plot, red, black, blue, green, magenta, purple and
orange correspond to $\cm$ =$10^{-3}$, 0.1, 0.2, 0.3, 0.6, 1 and 100
respectively. $\s$ is in units of $(16\pi G_3)^{-1}$ and $\o$ is in units of $2r_H^2/L^3$.}
}
\par
 In figure (\ref{Cvw}), we show the dependence of conductivity on frequency for various fixed values of $\cm$ for the single and equal charged case. Here, the behaviour of the curves are in contrast with each other in two cases. While for the single charged case, we find the curves fit well with the expression $\sim a(\cm) +\f{b(\cm)}{c(\cm)+\o ^2} $  for some $\o$ independent functions $a(\cm),~ b(\cm)$ and $c(\cm)$ of $\cm$. On the right, we see that the conductivity increases for intermediate values of $\cm $ as $\o$ is increased.

\subsection{The relation to the M2-brane theory}

It has been observed that the thermodynamic properties of the near horizon 
geometry   of M2-branes is  similar to  that of the D1-branes
\cite{Peet:1998wn,Mateos:2007vn}.  We now recall the thermodynamic 
properties of uncharged M2-branes and compare them to uncharged 
D1-branes. These properties were obtained from \cite{Harmark:1999xt}. 
The near horizon geometry of M2-branes is $AdS_4$ times $S^7$, let the radius of $S^7$ be $L'$ and 
the Newton's constant in $4$ dimensions be $G_4$.
The thermodynamic properties of non-extremal uncharged D1-branes and non-extremal uncharged M2-branes with 
non-extremal parameter $r_0$ is given by:
\begin{center}
\begin{tabular}{|l|l|l|}
\hline
  & D1-branes & M2-branes \\ 
\hline \hline&&\\
$s$ &  $\frac{1}{4G_3} \frac{r_0^4}{L^4}$ & $\frac{1}{4G_4} \frac{r_0^4}{L^{\prime4}}$ \\
&&\\
\hline&&\\
$T$ & $\frac{3}{2\pi} \frac{r_0^2}{L^3}$ &  $\frac{3}{2\pi} \frac{r_0^2}{L^{\prime 3} }$ \\
&&\\
\hline&&\\
$\epsilon$ & $\frac{1}{4\pi G_3} \frac{r_0^6}{L^7}$  &   $\frac{1}{4\pi G_4}
 \frac{r_0^6}{L^{\prime 7}}$ \\
&&\\
\hline&&\\
$ p = -f $ &  $\frac{1}{2} \epsilon$ &  $\frac{1}{2} \epsilon$ \\
&&\\
\hline 
\end{tabular}\\
\vspace{.5cm}
{\bf Table 2. } Thermodynamics of uncharged D1-branes and M2-branes.
\end{center}
From the equation of state $p = \frac{\epsilon}{2}$, it seems that the 
non-conformal D1-brane theory behaves as though it is a conformal theory in $2+1$ dimensions. 

This similarity of thermodynamic properties of uncharged D1-branes and M2-branes
was also seen to extend to the transport properties. 
In \cite{David:2009np}, it was noted that the bulk viscosity to entropy density of non-extremal D1-branes is given by 
$1/4\pi$.  This fact  was explained by the observation in \cite{Kanitscheider:2009as}. 
Consider  conformal hydrodynamics  of a charged fluid in $2+1$ dimensions
\footnote{ \cite{Kanitscheider:2009as} considered 
the case of the uncharged fluid and obtained a general 
relation between  conformal hydrodynamics  between $2\sigma$ dimensions
and non-conformal hydrodynamics in $d$ dimensions.  The case when $\sigma =3/2, d=2$
corresponds to the relation between D1-branes and M2-branes. In general, the 
relation found in \cite{Kanitscheider:2009as} relates conformal hydrodynamics 
in fractional dimensions to non-conformal hydrodynamics in integer dimensions.}.
 The stress tensor  and the current are given by
\begin{eqnarray}
\tilde T^{ab} &=& \tilde \epsilon u^{a} u^b + \tilde p ( \eta^{ab} + u^a u^b ) 
 - 2 \eta \sigma^{ab}, \\ \nonumber
\tilde j^a &=& \tilde \rho u^a  - \tilde \sigma T( \eta^{ab} + u^au^b)  \partial_b\left( \frac{\mu}{T} \right), 
\end{eqnarray}
where $a, b \in \{0, 1, 2 \}$, $\eta_{ab}$ is the Minkowski metric in 
$2+1$ dimensions and 
\begin{equation}
 \sigma_{ab} = P_a^c P_b^d \partial_{( c} u_{d)} - \frac{1}{2} P_{ab} \partial\cdot u , 
\qquad P_{ab} = \eta_{ab} + u_{a} u_b. 
\end{equation}
Let us now dimensionally reduce these equations with the ansatz
$u^a = ( u^\mu, 0)$ where $\mu \in\{0, 1\}$ along with the assumption that
there is no dependence along the  direction $2$ for any thermodynamic variable. 
Then the non-trivial components of the stress tensor and 
the current can be written as
\begin{eqnarray}
\tilde T^{\mu\nu} &=&  ( \tilde \epsilon + \tilde p) u^{\mu} u^\nu + \tilde  p g^{\mu\nu} 
-2 \eta \tilde \sigma ^{\mu\nu} - \eta P^{\mu\nu} \partial\cdot u, \\ \nonumber
\tilde T^{2\mu} &=& \tilde T^{22} =0, \\ \nonumber
\tilde j^\mu &=&  \tilde \rho u^\mu  - 
\tilde \sigma T( g^{\mu\nu} + u^\mu u^\nu )  \partial_\nu\left( \frac{\mu}{T} \right), 
\\ \nonumber
j^2 &=&0, 
\end{eqnarray}
where 
\begin{equation}
 \tilde \sigma_{\mu\nu} = P_{\mu}^\rho P_{\nu}^\sigma 
\partial_{( \rho}u_{\sigma)} -  P_{\mu\nu}\partial\cdot u =0.
\end{equation}
To show the above expression vanishes, one can explicitly evaluate the 
components or else use the fact that it is a traceless
symmetric tensor in $1+1$ dimensions  and is orthogonal to the 
velocity vector $u^\mu$. 
Thus the stress tensor and the charge current in $1+1$ 
dimensions is given by 
\begin{eqnarray}
T^{\mu\nu} &=& R \tilde T^{\mu\nu} = 
( R\tilde \epsilon + R\tilde  p) u^{\mu} u^\nu + R\tilde  p \eta^{\mu\nu} 
 - R\eta P^{\mu\nu} \partial\cdot u, \\ \nonumber
j^\mu &=& R\tilde j^\mu = R\tilde \rho u^\mu  - 
 R\tilde \sigma T( \eta^{\mu\nu} + u^\mu u^\nu )  \partial_\nu\left( \frac{\mu}{T} \right), 
\end{eqnarray}
where $R$ is the radius of compactification.  On comparing this form of the stress tensor to that given in  (\ref{fluidstress}) we see that 
we can identify
\begin{equation}
\label{lowdim}
 \epsilon = R\tilde\epsilon, \qquad p = R\tilde \epsilon, \qquad \sigma = R\tilde \sigma,\qquad
\zeta = R\eta. 
\end{equation}
The entropy density $\tilde s$ in $2+1$ dimensions is related to the 
entropy density in $1+1$ dimensions by 
\begin{equation}
 s = R\tilde s. 
\end{equation}
From this, we can conclude that
for a fluid dynamics  in $1+1$ dimensions,    which is related by compactification 
on a circle of radius $R$ to  conformal hydrodynamics in $2+1$ dimensions, 
 the relation  
\begin{equation}
 p = \frac{\epsilon}{2} 
\end{equation}
will continue to hold true due to (\ref{lowdim}).  Furthermore,
we have
\begin{equation}
\label{dimred1}
 \frac{\zeta}{s} = \frac{\eta}{\tilde s}. 
\end{equation}
Thus the ratio of bulk viscosity to entropy density  in $1+1$ dimensions is identical to the ratio of  shear viscosity to entropy density of the conformal $2+1$ hydrodynamics. 

In \cite{Kanitscheider:2009as}, it was shown that the
equations of  gravity fluctuations for the uncharged D1-brane which determine the hydrodynamical 
transport coefficients  is a dimensional reduction of the gravity fluctuations of the 
uncharged M2-brane background.  This fact and (\ref{dimred1}) 
explains the reason why the ratio of bulk viscosity to entropy density 
for the D1-brane is given by $1/4\pi$.  It also explains the fact that 
speed of sound for the D1-brane theory is same as that of the M2-brane theory.
One expects this argument to go through for the charged 
D1-branes and this  is  the reason we observe that the speed of sound 
is $1/\sqrt{2}$ and 
the bulk viscosity to entropy density is $1/4\pi$. 
As an evidence for this argument, 
 we will now show  that the 3 dimensional  truncated action 
given in (\ref{threed})   which  supports the equal charged D1-brane solution can be 
obtained by dimensional reduction of the following $4$ dimensional action. 
\begin{equation}
\label{ads4}
  S = \frac{1}{16\pi G_4} \int d^{4}x \sqrt{-g_{4}}\lf (R_{4} + \frac{6}{L^{\prime 2} }  -
 L^{\prime 2} F^{\mu\nu}F_{\mu\nu}\ri ), 
\end{equation}
where $g_4$ and $R_4$ are the $4$ dimensional metric and the  Ricci curvature 
respectively.  $G_4$ in the four dimensional Newton's constant and $L'$ is the 
radius of $AdS_4$. 
This is the action which admits the solution of the equal charged M2-brane. 
Note that the near horizon geometry of the  equal charged M2-brane is just a 
Reissner-Nordstr\"{o}m
black hole in $AdS_4$.  
We address the equal charged case since the single charged M2-brane has 
not been studied in the literature. 
We now compactify the action in (\ref{ads4}) using the following ansatz
\begin{eqnarray}
 ds^2 &=& ds^2_{(2+1)} + e^{-\frac{4}{3} \phi } dy^2, \\ \nonumber
A_y &=& 0.
\end{eqnarray}
As usual, all fields do not have any dependence on the compact direction $y$. 
Substituting this ansatz in the action (\ref{ads4}), we obtain 
\begin{equation}
 S = \frac{2\pi R_y}{16\pi G_4} 
  \int d^3 x \sqrt{- \tilde g_3} e^{ -\frac{2}{3} \phi} \lf ( \tilde R_3 + \frac{6}{L^{\prime 2} }
-  L^{\prime 2}  F^{\mu\nu} F_{\mu\nu}\ri ). 
\end{equation}
To bring the action in the Einstein form, we perform the following re-definition
\begin{equation} 
 \tilde g_{\mu\nu} = e^{\frac{4}{3} \phi} g_{\mu\nu}. 
\end{equation}
We then obtain 
\begin{eqnarray}
\label{dimred}
S &=& \frac{2\pi R_y}{16\pi G_4}
  \int d^{3}x \sqrt{-g_{3}}\lf (R_{3}-\frac{8}{9}(\partial\phi)^2+
\frac{6 }{L^{\prime 2}} e^{\frac{4}{3}\phi}
 -L^{\prime 2} e^{\frac{-4}{3}\phi}F^{\mu\nu}F_{\mu\nu}\ri ). 
\end{eqnarray}
Now  comparing (\ref{threed}) and the above action, we see that they are the same
on identifying
\begin{equation}
 \Phi=  \exp\lf ( \frac{2}{3} \phi\ri ) , \qquad L' =  \frac{L}{2} , \qquad  
  A_\mu \rightarrow \frac{A_\mu}{ L'}. 
\end{equation}
This observation indicates that the 
supergravity  fluctuations which determine the transport coefficients 
of the equal charged  D1-brane theory can be obtained by dimensional reduction of 
the fluctuations which determine the transport coefficient of the equal charged
M2-brane theory. As a result, the transport coefficients of the M2-brane theory 
is related to that of the D1-brane theory. 

Finally, we mention that from (\ref{lowdim}) the conductivity of the M2-brane theory is
related to that of the M2-brane theory. The conductivity of the equal charged
M2-brane theory has been evaluated in \cite{Hartnoll:2007ip} 
\footnote{see equation (83) of \cite{Hartnoll:2007ip} and identify
$q^2$ as $k$} and is given by 
\begin{equation}
\sigma_{\rm{M2}} = \frac{1}{16\pi G_4} \frac{(3-k)^2}{9 ( 1+k)}.
\end{equation}
Note that apart from the dimensions set by $G_4$, the dependence of the 
conductivity is identical to that of the equal charged D1-brane theory given in 
(\ref{equalc}). In table 1, we have compared the transport properties of the
equal charged M2-brane and the equal charged D1-brane.

\section{Conclusions}

In this paper, we have studied the transport properties of 
the $1+1$ dimensional $SU(N)$ gauge theory with 16 supercharges of
the D1-branes at finite chemical potential in the framework of the 
gauge/gravity duality.
We evaluated the bulk viscosity, electrical conductivity , thermal
conductivity, the charge and sound diffusivity for two cases. 
One in which
the chemical potential conjugate to one of the $U(1)$ R-charges is turned on 
and another in which equal charges conjugate to all the $4$ Cartans of 
the $SO(8)$ R-symmetry are turned on. 
In both the situations, we find that the ratio of bulk viscosity to the entropy
density is independent of the chemical potential and is equal to $1/4\pi$. 
We showed that for the single charged D1-brane theory, the charge dissipative mode
becomes unstable  and 
the transport properties exhibit critical behaviour at the boundary of 
thermodynamic instability. We also demonstrated that the shear viscosity and 
thermal conductivity satisfy a relationship similar to the Wiedemann-Franz law. 
We have observed that the transport coefficients of the D1-branes theory 
is same as that of the M2-brane theory apart from  an overall 
normalization which determines the dimensions and suggests a plausible 
reason for this behaviour.
The summary of the transport coefficients obtained in this paper and 
their comparison with the transport coefficients of the M2-brane theory is given in 
table 1. 
 A technical result of our analysis is the following:
we reduced the problem of solving the second order differential equation
which determines the transport coefficient to a first order non-linear
differential equation. This equation  
governs the radial evolution of the transport coefficient. We were
able to solve these equations analytically for the transport coefficients of 
interest in this paper.

A possible extension of this work is to compute the transport coefficients
when all the $4$ chemical potentials corresponding to the 
$4$ Cartans of the $SO(8)$ R-symmetry are turned on. 
This would provide a complete knowledge of the transport coefficients of the 
D1-brane gauge theory. 
It will also be interesting to understand the thermal stability of the full system 
with all the R-charges turned on. 
Another direction is to understand the connection of the 
D1-brane theory with that of the M2-brane theory better. This would involve
an analysis similar to \cite{Kanitscheider:2008kd}. We need to show
that the  hydrodynamic fluctuations
in gravity
 which determine the transport for the charged M2-brane and D1-brane are related by compactification.  From the point of the view of the theories of the M2-branes
 and D1-branes,  it is interesting to note that unlike the 
 presently unknown theory of the M2-branes, 
 the theory of the D1-brane is a regular gauge theory in $1+1$ dimensions. 
 We have seen that  the D1-brane gauge theory  provides physical information regarding 
 the M2-brane theory. It is worthwhile to explore and utilize   this fact 
 to understand the M2-brane theory further. 
 
 $1+1$ relativistic hydrodynamics occurs in the  short time description of the plasma
 formed after highly relativistic collisions \cite{Bjorken:1982qr}. 
 The equation of state of this 
 plasma does not obey $p = \epsilon/2$, however it will be interesting to see
 if the transport properties of this plasma show the behaviour seen here.
 Another area where relativistic $1+1$ hydrodynamics could be 
 important is in carbon nano-tubes and graphene nano-ribbons. 
 These materials can be described as  a graphene layer rolled up
 and a graphene layer whose linear dimensions is much larger than
 that of its width respectively 
 \cite{PhysRevLett.78.1932,PhysRevB.73.235411}. These systems are relativistic 
 since they are  obtained by 
 a dimensional reduction of  $2+1$ dimensional graphene which is described
 by a massless Dirac equation. It will be interesting to compare the 
 transport properties of these materials with that of the field theory 
 studied here.
 The system we study here has a gap set by the Yang-Mills coupling.
Hydrodynamics of other  1+1 dimensional systems with a gap
have been studied in \cite{PhysRevB.57.8307, PhysRevB.59.9285}
\footnote{We thank Subir Sachdev for bringing these
references to our notice.}.
 Even though we have analysed only a bosonic system, we can think of it describing a 1+1 dimensional condensed matter system or a quasi 1+1 dimensional system made up of strongly interacting bosonic quasiparticles which are themselves made up of elementary electrons, just like Cooper pairs. It would be interesting to evaluate an effective Lagrangian for such quasiparticles from the action of the gauge theory dual to our gravity system and then compare it to effective action for one dimensional effective condensed matter systems like Luttinger liquids. A curious observation is that our plots for conductivity vs temperature and frequency for equal charged case qualitatively looks similar to a system of carbon nanotubes-polyepoxy composites \cite{barrau}.

\acknowledgments
We wish to thank Pallab Basu for useful discussions and 
help in setting up the initial numerical programs developed for this paper.
We wish to thank Sean Hartnoll, Chris Herzog and Subir Sachdev
for several useful comments on an earlier version of the manuscript.
We especially thank Subir Sachdev for comments which helped us to
present our results to a wider audience.  
We also wish to thank Rajesh Gopakumar, Shiraz Minwalla and 
Shiroman Prakash for discussions. 
We thank the ICTS,  
TIFR for organizing a   stimulating discussion meeting on 
current topics in string theory at CHEP, IISc  during which part of this work 
was completed.

\appendix

\section{Consistent truncation to 3 dimensions}

We first show that
the solution (\ref{finaltruncsol}) in 3 dimensions   is a consistent  truncation of the 
spinning D1-brane solution 
in 10 dimensions given in (\ref{spinmetric}). 
For this, we use the results of \cite{Cvetic:2000dm} who gave the most general 
ansatz for the consistent Kaluza-Klein reduction of a 10 dimensional 
solution on the seven sphere \footnote{See section 5. of \cite{Cvetic:2000dm}. }. 
The ansatz is as follows:
\bea
\label{CLP}
ds_{10}^2&=&Y^{\f{1}{8}}\lf [\D _C^{\f{3}{4}}ds_3^2+g^{-2}\D _C^{-\f{1}{4}}T_{ij}^{-1}\cD{\m}^i\cD \m ^j \ri ], \nn\\
e^{-2\phi}&=&\D _C^{-1}Y^{1/2}, \nn\\
\hat{F}_{(3)}&=&F^1+F^2+F^3. 
\eea 
where \footnote{Note that the sign of $F^1$ here is negative of that in \cite{Cvetic:2000dm}, this is 
a result of  a different convention for the volume form $\ep_3$.}
\bea
F^1&=&gU\ep _3, \nn\\
F^2&=&g^{-1}T_{ij}^{-1}* \cD T_{jk}\w (\m ^k \cD \m ^i), \nn\\
F^3&=&-\f{1}{2g^2}T_{ik}^{-1}T_{jl}^{-1}* F^{ij}_{(2)}\w \cD \m ^k\w \cD \m ^l, \nn\\
\cD\m ^i&=&d\m ^i+gA^{ij}\m ^j, \nn\\
\cD T_{ij}&=&dT_{ij}+gA^{ik}T_{kj}+gA^{jk}T_{ki}, \nn\\
F_{(2)}^{ij}&=&dA^{ij}+gA^{ik}\w A^{kj}, 
\eea
and
\bea
\m ^i\m ^i=1{\hspace{.7 cm}}\D _C=T_{ij}\m ^i\m ^j{\hspace{.7 cm}}U=2T_{ik}T_{jk}\m ^i\m ^j-\D _{C}T_{ii}{\hspace{.7 cm}}Y={\tr{ det}}(T_{ij}). 
\eea
 $*$ is the Hodge dual in the three dimensions. The $\mu$'s are defined as follows, 
\bea
\label{munu}
\nu _1=\sin\t, &\qquad&
\nu _2=\cos \t\sin\psi_1,\nn\\
\nu_3=\cos\t\cos\psi_1\sin\psi_2, &\qquad&
\nu_4=\cos\t\cos\psi_1\cos\psi_2\nn,\\
\mu _{2a-1}=\nu_a \sin\phi _a, &\qquad& 
\mu_{2a}=\nu_a\cos\phi _a.
\eea
Here, $a=1, \cdots 4$ and $i, j = 1, \cdots 8$. 
Then \cite{Cvetic:2000dm}  shows that on substituting the above ansatz in to the 
 ten dimensional equations of motion, there is a consistent reduction to the equations
 of motion for the three dimensional fields. The equations of motion for the 
 three dimensional fields can be derived from the following three-dimensional 
 Lagrangian:
\bea
\label{eff3D}
\cL &=&R* 1-\f{1}{32}Y^{-2}* dY\w dY-\f{1}{4}{\ti T}_{ij}^{-1}* \cD {\ti T}_{jk}\w {\ti T}^{-1}_{kl}\w\cD{\ti T}_{li}\nn\\&&
-\f{1}{4}Y^{-1/4}{\ti T}^{-1}_{ik}{\ti T}_{jl}^{-1}* F^{ij}_{(2)}\w F^{kl}_{(2)}-\f{g^2}{2}Y^{1/4}\{2\ti{T}_{ij}\ti{T}_{ij}-(\ti{T}_{ii})^2\}* 1,
\eea
where ${\ti T}_{ij}=Y^{-1/8}T_{ij}$. 

\vspace{.5cm}
\noindent
{\bf Single charged D1-brane}
\vspace{.5cm}

We now show that the spinning D1-brane solution in 10 dimensions given in 
(\ref{spinmetric}) can be written in the form given in (\ref{CLP}). For this, 
we choose
\bea
\label{sinanst}
g&=&\f{1}{L}, \qquad
A^{12}=-\f{r_0^3l}{L^2r^2H}dt, \nn\\
H&=&1+\f{l^2}{r^2}, \qquad
T_{ij}=X_{(i)}\d _{ij}, \nn\\
X_{(i)}&=&\f{L^2}{r^2H} \quad (i=1,2), \qquad
X_{(i)}=\f{L^2}{r^2} \quad (i\neq 1,2), \nn\\
\ep _{tzr}&=&1. 
\eea
For convenience, we also write down the following
\begin{eqnarray}
\D _C&=&\f{L^8}{r^8}\f{1}{HH_1}, \qquad
H_1=\f{L^6}{\D r^6}, \nn\\
{\D}&=&1+\f{l^2}{r^2}\cos ^2\t, \qquad
Y=\f{L^{16}}{r^{16}}\f{1}{H^2}\nn\\
U&=&-2\f{L^4}{r^4H}\lf (3+\f{2l^2}{r^2}\cos ^2\t\ri ). 
\end{eqnarray}
This results in the metric
\bea
\label{apspin}
ds^2 &=&  H_1^{-3/4} (  -f dt^2 + dz^2) + H_1^{1/4} \lf( \f{1}{\ti h} dr^2  + r^2 ( \D d\t ^2 + \ti \D \sin^2 \t d\phi^2 + \cos^2 \t d\O _5^2)
 \ri) \nn\\ 
&& -2 H_1^{-3/4} \f{ L^3 r_0^3}{\D r^6} l \sin^2 \t dt d\phi, 
\eea
which is same as that given by equation (\ref{spinmetric}).
Note that the exponent of dilaton is negative of that given in main text. This is just due to difference in conventions between \cite{Harmark:1999xt} and \cite{Cvetic:2000dm}.
The dilaton and the three form are given by
\bea
e^{\p}&=&H_1^{-1/2}, \nn\\
\hat{F}_{(3)}&=&-2\f{r^5}{L^6}\lf (3+\f{2l^2}{r^2}\cos ^2\t\ri )dt\w dz\w dr+
2\sin\t\cos\t\f{l^2r^4}{L^6}dt\w dz\w d\t\nn\\&&-2\sin\t\cos\t\f{r_0^3l}{L^3}dz\w d\t\w d\phi.
\eea
This also agrees with the expression for the dilaton modulo the 
sign and the two form gauge potential given in  (\ref{spinmetric}). 
One can also check that by reading out the three dimensional metric by comparing  (\ref{CLP}) to (\ref{apspin})  , we obtain  the three dimensional truncated solution in 
(\ref{finaltruncsol}).
We can now proceed to obtain the three dimensional Lagrangian for the 
Kaluza-Klein ansatz in (\ref{sinanst}). We first 
define the 
 scalars $Z_1=Y^{-1/8}X_1=Y^{-1/8}X_2$ and $Z_2=Y^{-1/8}X_j$ for $j\neq 1, 2$. 
 Then the Lagrangian in (\ref{eff3D}) reduces to
\bea
\cL&=&\sqrt{-g}\lf [R-\f{1}{32Y^2}\pa _{\m}Y\pa ^{\m}Y
-\f{1}{2}\lf (\f{1}{Z_1^2}\pa _{\m}Z_1\pa ^{\m}Z_1+\f{3}{Z_2^2}\pa _{\m}Z_2\pa ^{\m}Z_2\ri )
\ri .\nn\\&&\lf .-\f{1}{4Y^{1/4}}\f{1}{Z_1^2}F_{\m\n}F^{\m\n}+\f{12}{L^2}Y^{1/4}Z_2(Z_1+Z_2)\ri ]. 
\eea
On identifying 
\bea
Z_1&=&\Psi^{-3/4}, \qquad
Z_2=\Psi^{1/4}, \qquad  Y^{1/4} = \frac{e^{\frac{4}{3} \phi}}{\Psi^{1/2}}, 
\eea
the above action reduces to the one given by equation (\ref{truncact}).

\vspace{.5cm}
\noindent
{ \bf Equal charged D1-brane}
\vspace{.5cm}

We now wish to obtain the truncated 3 dimensional solution as well as the 
action when one turns on equal charges along the $4$ Cartans of the $SO(8)$ 
R-symmetry. 
We start with the 10 dimensional D1-brane solution with equal spins along the 
$4$ Cartan's. This is given by \cite{Harmark:1999xt}.
\bea
ds^2&=&H_2^{-3/4}(-fdt^2+dz^2)+H_2^{1/4}\lf (\frac{dr^2}{h\bar{f}}+\L_{\a\b}d\eta ^{\a}d\eta ^{\b}\ri )\nn\\
&&-2\f{H_2^{-3/4}}{h^3}\f{L^3r_0^3}{r^6}l\sum_{i=1}^{4}\nu_i^2dtd\phi _i, \nn\\
A_2&=&-\lf(H_2^{-1}dt+\f{r_0^3}{L^3}l\sum_{i=1}^4 \nu _i^2d\phi _i\ri )\w dz, \nn\\
e^{\p}&=&H_2^{1/2}, \qquad
H_2=\f{L^6}{r^6h^3}, \nn\\
h&=&1+\f{l^2}{r^2}, \qquad
f=1-\f{r_0^6}{h^3r^6}, \qquad
\bar{f}=1-\f{r_0^6}{h^4r^6}\nn\\
\L_{\a\b}d\eta ^{\a}d\eta ^{\b}&=&r^2h[d\t ^2+\cos ^2\t d\psi_1^2+\cos ^2\t\cos^2\psi_1d\psi _2^2+\sin ^2\t d\phi_1^2\nn\\
&&+\cos ^2\t\sin ^2\psi_1 d\phi_2^2+\cos^2\t\cos^2\psi_1\sin^2\psi_2d\phi_3^2\nn\\&&
\label{HOsoln}
+\cos^2\t\cos^2\psi_1\cos^2\psi_2d\phi _4^2]. 
\eea
We will now compare  the 10 dimensional solution with the 
form of the Kaluza-Klein ansatz given in  (\ref{CLP}). 
For this, we first assume that the three dimensional metric is of the form
\be
ds_3^2=Z\lf[-\hat{f}dt^2+dz^2+\f{H_2}{h\bar{f}}dr^2\ri ], 
\ee
 and 
 \begin{equation}
 T_{ij}=\P\d _{ij}, \qquad A^{ij}=a(r)\s^{ij}  dt, 
 \end{equation}
  where $\s ^{2a-1,2a}=-\s^{2a,2a-1}=1$ and zero otherwise.  $\mu_i$
  are given in (\ref{munu}). 
  With this ansatz,  the gauge field is given as 
\bea
F_{3}&=&gUZ\lf (\f{L^3}{r^3h^2}\sqrt{\f{Z\hat{f}}{\bar{f}}}\ri )dt\w dz\w dr\nn\\
&&+\f{a'}{g^2\P ^2}\lf (\f{r^3h^2}{L^3}\sqrt{\f{\bar{f}}{Z\hat{f}}}\ri )dz\w d\p _a\w \nu _ad\nu _a.
\eea
Comparing with field strength of solution in  (\ref{HOsoln}), we get
\bea
\label{mina}
\f{L^3}{r^3h^2}\sqrt{\f{Z\hat{f}}{\bar{f}}}=\f{1}{g\P ^2 r h ZH_2}, \qquad
\f{a'}{g}Z=2\f{r_0^3l}{L^9}r^5h^2.
\eea
By comparing the metric in (\ref{CLP}) and the spinning D1-branes solution 
  (\ref{HOsoln}), we get
\bea
\label{mina2}
\P ^{7/4}Z\hat{f}-\f{a^2}{\P ^{1/4}}=H_2^{-3/4}f, &\qquad&
\P ^{7/4}Z=H_2^{-3/4}, \nn\\
g^2\P ^{1/4}r^2h=H_2^{-1/4}, &\qquad&
\f{a}{g\P ^{1/4}}=-\f{H_2^{-3/4}}{h^3}\f{l L^3r_0^3}{r^6}. 
\eea
A solution to the equations in (\ref{mina}) and (\ref{mina2})  is given by
\bea
g=L^{-1}, &\qquad& 
\P^{-1}=g^2r^2h, \nn\\
a=-\f{g^2r_0^3l}{r^2h}, &\quad& 
Z=(g^2r^2h)^4, \nn\\
\hat{f}= \bar{f}.  &\quad& 
\eea
Now using these equations, the effective 3-dimensional action as given by equation (\ref{eff3D}) reduces to
\be
\label{threed}
{\cal{L}}_3=\sqrt{-g}\lf[R-\f{2}{\P ^2}\pa _{\mu}\P\pa ^{\mu}\P -\f{1}{\P ^2}F_{\m\n}F^{\m\n}+24\f{\P ^2}{L^2} \ri ], 
\ee 
with the 3 dimensional  solution
\bea
\label{metEC}
ds_3^2&=&\f{h^4}{L^8u^4}\lf [-fdt^2+dz^2+\f{L^6}{4h^4f}du^2\ri ], \nn\\
h&=&1+k u, \nn\\
A&=& -\f{r_0^3l}{L^2h} u dt, \nn\\
f&=&1-\f{r_0^6u^3}{h^4}, \nn\\
\P &=&\f{L^2u}{h}. 
\eea
We have changed the radial variable to $u=\f{r_H^2}{r^2}$, where $r_H$ is the radius of the horizon. Here, $k=\frac{l^2}{r_H^2}$. We have divided by $r_H$ to turn some quantities like $L,\; r_0$ above to be dimensionless.
The variable $f$ above is same as $\bar{f}$ in earlier part of the analysis. 
The radius of the 
seven sphere  is $L$. Parameter $r_0$ is related to $k$ and 
the radius of the horizon as 
\be
\label{r0EC}
r_0^6=(1+k)^4 r_H^6. 
\ee
Note that there is no extra scalar in this case. 
The general compactification given in \cite{Cvetic:2000dm} contains 36 scalars, one singlet under SO(8) and the rest which transforms  as {\bf 35}. 
In this equal charged case, we turn on only the singlet which is the dilaton. 
 In the single  charged case, 
  one more scalar $\Psi$  is turned on and this  explicitly breaks  the $SO(8)$
   symmetry.

\section{Transport coefficients  for the equal charged D1-brane}

In this part of the appendix, we evaluate the conductivity and bulk viscosity for the 
equal charged D1-brane.  We will be brief here since we have provided the details 
for the single charged D1-brane in the main text. 

\noindent
We first provide a table listing the thermodynamic properties of the equal charged D1-brane

\begin{center}
\begin{tabular}{|l|l|}
	\hline
&\\
Hawking Temperature($T$)   & $\f{r_0^3}{4\pi L^3 r_H} (\f{6-2k}{1+k})$    \\
&\\
Entropy Density($s$)  & $\frac{1}{ 4 G_3} \frac{ r_0^3 r_H}{L^4}$     \\
&\\
Energy Density($\epsilon$)   & $\frac{1}{4\pi G_3} \frac{ r_0^6}{L^7}$   \\ 
&\\
Pressure $\equiv$(-free energy density($f$))  & $\frac{1}{8\pi G_3}\f{r_0^6}{L^7}=\f{\epsilon}{2}$    \\
&\\
Charge Density($\rho$)   & $\f{r_{0}^3l}{8\pi G_{3}L^5}$   \\
&\\
Chemical Potential($\mu$)   & $\f{lr_{H}(1+k)}{L^2}$   \\
&\\
	\hline
\end{tabular}\\
\vspace{.5cm}
{\bf Table 3.}  Thermodynamic properties of the equal charged D1-brane
\end{center}
In evaluating the above thermodynamic quantities, we have used the relation in (\ref{r0EC}). 
Note that the   Hawking temperature of this black hole  is given by
\begin{equation}
T =\f{r_0^3}{4\pi L^3 r_H}\left(\f{6-2k}{1+k}\right).
\end{equation}
From this expression, we see that the black hole is stable only for
\begin{equation}
k<3. 
\end{equation}
We can also examine the Hessian to see if the equal charged solution 
admits the thermodynamic instability seen in the case of the single charged 
solution. Using the expression of the Hessian given in (\ref{convhes}), we obtain
\begin{equation}
 H_s  = \f{8 G_{3}^2 L^4(k+3)}{r_{H}^4(1+k)^2}. 
\end{equation}
Since $k\geq0$, the Hessian for this case
 is always positive and therefore this solution does not exhibit the usual 
 thermodynamic instability. Thus the range of the allowed values of $k$ is 
 $0<k <3$. This is the same range found for the case of  equal charged M2-branes
 \cite{Hartnoll:2007ai}. Using the expressions for the thermodynamic variables given in 
 table 3, we can evaluate the relationship between the 
 charge diffusion constant and the conductivity from the formula in (\ref{expcdc}). 
 It is given by
 \begin{equation}
 D_c = \sigma (16\pi G_3) \frac{ 3 (k+3)}{2 r_H^2 ( 3-k)^2}. 
 \end{equation}

\vspace{.5cm}\noindent
{\bf Hydrodynamic modes from gravity}
\vspace{.5cm}

To obtain the two hydrodynamic modes of the charged fluid from gravity, 
we  analyze linearized wave like perturbations in the background of the equal charged D1 brane solution given in  (\ref{metEC}). It is a solution of the action given in (\ref{threed}).  The perturbations are defined as follows: 
\bea
g_{tt}\ra g_{tt}^0(1+H_{tt}), &\qquad&
g_{tz}\ra g_{zz}^0H_{tz}, \nn\\
g_{zz}\ra g_{zz}^0(1+H_{zz}), &\qquad& 
A_t\ra A_t^0+\f{lr_0^3}{L^2}B_t, \nn\\
A_z\ra \f{lr_0^3}{L^2}B_z, &\qquad& 
\P\ra \P ^0+L^2\vp. 
\eea
where the superscript `$0$' refers to the background values. 
Due to translational invariance along the $t$ and the $z$ directions, we can assume 
that the dependence of the perturbations along 
these  directions  is of the form
as $\sim\exp[\f{2i}{L^3}(-\o t+q z)]$. Note that here we  will be using the 
dimensionless variables defined in (\ref{dimension}). 

To write the gauge invariant modes, we first introduce the following functions
\bea
V&=&q^2(4h^3-r_0^6u^3)-4\o ^2h^3,\nn\\
\a &=&q^2 f-\o ^2.
\eea
The  two gauge invariant variables which are invariant both under
 diffeomorphism as well as $U(1)$ gauge 
transformations are given by
\bea\label{ginveq}
Z_P&=&-q^2 f H_{tt}+2\o q H_{tz}+\o ^2 H_{zz}+\f{V}{h^2u}\vp ,\nn\\
G_P&=&q B_t+\o B_z+q\vp .
\eea 
These gauge invariant variables satisfy the following equations of motion
\bea
\label{equaleqn}
V\a f u Z''_P&-& 
 2 h^2 (q^2 - \o^2) \{q^2 (h^2 + 5 h - 12) - 
    2 h \o^2\} (1 - f)Z'_P\nn\\&&+
 q^2 h^2 (1 - f)^2 \{q^2 (2 h^2 + 2 h - 8) - w^2 (h^2 + 6 h - 8) + 
    q^2 f h (3 h - 8)\}Z'_P\nn\\&&
+8 (h - 2) h^2 (q^2 - \o^2)^2Z'_P-2qkr_0^6\f{u^3V^2f}{h^7}G'_P\nn\\
&=&
-8 k q r_0^{12} \f{u^5 (4 - h)}{h^6} [3 q^4 f + 4 \o^4 - 
   q^2 \o^2 (4 + 3 f - h + f h)]G_P\nn\\&&
+\f{u\a}{fh}[4(q^2 - \o^2)^2 - q^2 (1 - f)  (h + 4) (q^2 - \o^2)\nn\\&& + 
  q^2 h (1 - f)^2 (3u^{-2} h^2 f (h - 4) +  q^2)]Z_P, \eea\bea 
G''_P&+&\f{q h}{V \a}\{q^2(2+h-hf)-2\o^2\}Z'_P+\lf [\f{2k}{h\a}(q^2-\o ^2)+
\f{\o ^2(4-h)(1-f)}{\a h fu}\ri ]G'_P\nn\\
&=&\f{3q  r_0^6u^2}{h^3fV}Z_P+\f{4kh(1-f)}{fuV\a}
\{6q^4f-q^2\o ^2(4+6f-h+fh)+4\o ^4\}G_P\nn\\&&
+\f{\a}{h^4f^2}G_P. 
\eea
Note that these equations decouple in the $q\rightarrow 0$ limit.


\vspace{.5cm}\noindent
{\bf{ Conductivity for the equal charged case}}
\vspace{.5cm}

In the limit $q=0$, the equation for the gauge invariant current mode is
\bea
\label{eqleqngge}
G''_P(u)+\f{(hf+h+2f-4)}{fhu}G'_P(u)+\f{\o^2}{h^4f^2}G_P(u)-\f{4k(1-f)}{h^2fu}G_P(u)=0.
\eea
We can remove the coefficient of $G_P$  proportional to $(1-f)$ by the following 
redefinition
\be
G_P=\f{4-h}{h}G. 
\ee
As we have seen for the case of the  single charged solution conductivity 
is essentially determined by the ratio 
\be
{\cal R}_{G_P} = \frac{1}{i\omega G_P} \frac{ d G_P}{du}, 
\ee
which in turn is determined by the ratio
 $g = \f{G_P'(u)}{i\o G_P(u)}$.
 Ingoing boundary conditions at the horizon for $G_P$ corresponds to the 
  following boundary condition on $g$
\be
\lim_{u\ra 1}g(u)=\f{1}{(3-k)(1+k)(1-u)}.
\ee
The equation of motion satisfied by $g$ is given by 
\bea
\label{eqlgeqn}
g'+i\o g^2 -\lf \{\f{(4-h)(1-f)}{fhu}+\f{2k}{(4-h)}\ri \}g-i\f{\o L^6}{4h^4f^2}=0. 
\eea 
The solution for the real part of $g$ in $\o \ra 0$ limit is 
\be
{\rm Re}\, g=\f{(3-k)^2}{(1+k)^2}\f{1}{(4-h)^2f}.
\ee
Just as  in the previous case for the single charged, the real part of DC conductivity here
 is proportional to   the value of ${\rm Re} {\cal R}_{G_P} $ at the boundary ($u=0$), which is given by
\be
\label{equalc}
{\tr Re}\, {\cal R}_{G_P}=\f{L^3}{2}\f{(3-k)^2}{9(1+k)^2}.
\ee
Note that here we have reinstated the factors of $L^3/2$ which we have absorbed in 
defining $\omega$. 
The imaginary part of ${\rm Re} {\cal R}_{G_P} $ is given by 
\be
{\rm Im}{\cal R}_{G_P}=
\f{L^3}{2}{\rm Im}\left[ \frac{d}{du}\ln \f{4-h}{i \o h} \right]_{u\ra 0}=\f{L^3}{2}\f{4k}{3\o}.
\ee
In Figure 3. we have compared these 
expressions with that determined by numerically solving the equation for $G_P$. 
We find that they agree to less than one part in $10^{-6}$. 
The DC conductivity is related to ${\rm Re} G_P$ by a proportionality constant
which can be determined  from the boundary effective action as was done
in the single charged case in section 5.3. This results in 
\begin{equation}
\sigma = \frac{1}{16\pi G_3} \frac{(3-k)^2}{9(1+k)^2}. 
\end{equation}

\vspace{.5cm}\noindent
{\bf{Bulk viscosity  for the equal charged case}}
\vspace{.5cm}

In this section, we calculate the bulk viscosity for the equal charged case and show that the ratio $\f{\zeta}{s}$ is constant.
In the $q=0$ limit, the equation for the sound mode decouples from the current mode. It turns out to be 
\bea
\label{Zeqn1}
Z''_P(u)+\f{2(h-2)}{fhu} Z'_P(u)-\f{(1-f)}{fu} Z'_P(u)+\f{\o^2 }{h^4f^2} Z_P(u)=0.
\eea
Let us define the ratio
\begin{equation}
g=  \f{Z_P'(u)}{i\o Z_P(u)}.
\end{equation}
Since the sound mode satisfies ingoing boundary condition at the horizon,
the  function $g$ should satisfy
\be
\lim_{u\ra 1}g=\f{1}{(3-k)(1+k)(1-u)}.
\ee
The appropriate solution for $g$ in $\o \ra 0$ limit is
 \be
{\rm Re}\, g= (1+k)^2\f{u^2}{h^4f}.
\ee
The bulk viscosity is proportional to  the real part of the following ratio
evaluated at the horizon:
\begin{equation}
{\cal R}_{Z_P} = \frac{1}{3i\omega u^2 Z_P } Z_{P}'(u). 
\end{equation}
We can evaluate this from the expression for ${\rm Re}\, g$ 
in the $\omega \ra 0$ limit which results in 
\begin{equation}
\left. {\rm Re}\, {\cal R}_{Z_P}\right|_{u\ra 0, \omega \ra 0}   
=\f {L^3}{2}\f{(1+k)^2}{3}.
\end{equation}
Here we have reinstated the factor of $L^3/2$  which we have absorbed in 
the definition of $\omega$. 
We have verified that the above expression using the numerical solution 
for the equation for $Z_P$ to one part in $10^{-9}$. This is shown 
in figure 4. 
Evaluating the proportionality constant relating the bulk viscosity to the 
ratio ${\rm Re}\, {\cal R}_{Z_P}$, we obtain
\begin{equation}
\label{bulkv}
\zeta = \frac{r_H^4}{16\pi G_3 L^4} ( 1+k)^2.
\end{equation}
The entropy density of the  equal charged solution (\ref{metEC}) is given by 
\bea
s=\f{r_{0}^3r_{H}}{4G_3L^4}.
\eea
Using (\ref{r0EC}) and the expression of the bulk viscosity (\ref{bulkv}), we get the ratio 
\be
\f{\zeta}{s}=\frac{1}{4\pi}. 
\ee

We can now evaluate the thermal conductivity of this solution using 
(\ref{thermcond}), this results in 
\begin{equation}
\kappa_T = \frac{r_H^2}{8LG_3}\frac{(1+k)(3-k)}{k}.
\end{equation}
It can also be verified that this system also satisfies the
Wiedemann-Franz like behaviour. 
\begin{equation}
\frac{\kappa_T\hat \mu^2}{\zeta T} = 4\pi^2. 
\end{equation}
The remaining transport coefficients $D_c$ and $D_s$ which are related to the 
conductivity and the bulk viscosity can be evaluated and are listed in 
table 1. In the end, we mention that we have verified that the  transport coefficients of the 
equal charged solution 
does not exhibit the critical behaviour seen in the 
case of the single charged solution in the domain of $0<k<3$. This is consistent 
with the fact that the Hessian does not show any sign of thermodynamic instability. 
When written in terms of 
$$\cm =\f{\mu}{2\pi L T} = \frac{\sqrt{k}}{3-k}, $$
the various transport coefficients are
\bea
\s &=&\f{1}{16\pi G_3}\lf [\f{1+2\sqrt{1+12\cm ^2}}{3(1+16\cm^2)}\ri ]^2,\nn\\
\z &=&\f{\pi L^2T^2}{4G_3}\lf [\f{1+\sqrt{1+12\cm ^2}}{6}\ri ]^2,\nn\\
\k _T &=&\f{\pi L^2 T}{4G_3}\lf (\f{1+6\cm ^2+\sqrt{1+12 \cm ^2}}{18\cm ^2}\ri ),\nn\\
D_c&=&\f{\sqrt{1+12\cm ^2}}{24\pi T }\f{(1+24 \cm ^2-\sqrt{1+12\cm ^2})}{\cm ^2 (1+16\cm^2)},\nn\\
D_s&=&\frac{1}{24\pi T} \lf (\frac{ 1+ 2\sqrt{1+12\cm^2}}{1+ 16\cm^2}\ri ).
\eea 

\bibliography{c-d1}
\bibliographystyle{JHEP}

\end{document}